\pdfoutput=1
\documentclass[aps,prd,reprint,twocolumn,showpacs,superscriptaddress,nofootinbib]{revtex4-1}  
\usepackage{tabularx}
\makeatletter
\def\hlinewd#1{
\noalign{\ifnum0=`}\fi\hrule \@height #1 
\futurelet\reserved@a\@xhline}
\makeatother
\usepackage{scrextend}
\usepackage{amsmath}
\usepackage{amssymb}
\usepackage{epsfig}
\usepackage{graphicx}
\usepackage{hyperref}
\usepackage{dcolumn}   
\usepackage{slashed}
\usepackage{color}
\usepackage{rotating}
\usepackage[margin=0.9in,a4paper]{geometry}
\usepackage[table,xcdraw,dvipsnames]{xcolor}
\usepackage[utf8]{inputenc}
\usepackage{colortbl}
\usepackage[normalem]{ulem}
\usepackage{mathrsfs} 
\usepackage[version=4]{mhchem}
\usepackage{cancel}

\definecolor{nicered}{rgb}{0.7,0.1,0.1}
\definecolor{nicegreen}{rgb}{0.07,0.35,0.07}
\definecolor{red}{rgb}{1.0, 0, 0}

\renewcommand{\(}{\left(}
\renewcommand{\)}{\right)}

\newcommand\g{\gamma}
\newcommand{\bdm}{\begin{displaymath}}
\newcommand{\edm}{\end{displaymath}}
\newcommand{\bea}{\begin{eqnarray}}
\newcommand{\eea}{\end{eqnarray}}

\renewcommand{\a}{\alpha}

\definecolor{nicered}{rgb}{0.7,0.1,0.1}
\definecolor{nicegreen}{rgb}{0.1,0.5,0.1}
\definecolor{red}{rgb}{1.0, 0, 0}
\definecolor{niceblue}{rgb}{0,0,0.8}
\definecolor{red}{rgb}{1.0, 0, 0}
\hypersetup{colorlinks,citecolor= nicegreen,linkcolor= nicered,urlcolor=nicered}

\def\eq#1{{Eq.~(\ref{#1})}}
\def\eqs#1#2{{Eqs.~(\ref{#1})--(\ref{#2})}}

\def\app#1{{Appendix~\ref{#1}}}

\def\vev#1{\left\langle #1\right\rangle}
\def\abs#1{\left| #1\right|}

\def\gsim{\raise0.3ex\hbox{$\;>$\kern-0.75em\raise-1.1ex\hbox{$\sim\;$}}}
\def\lsim{\raise0.3ex\hbox{$\;<$\kern-0.75em\raise-1.1ex\hbox{$\sim\;$}}}

\def\mb[#1]{\mathbf{#1}}

\renewcommand{\bar}{\overline}

\definecolor{LightCyan}{rgb}{0.88,1,1}
\definecolor{piggypink}{rgb}{0.99, 0.87, 0.9}
\definecolor{applegreen}{rgb}{0.55, 0.71, 0.0}
\definecolor{darkpastelgreen}{rgb}{0.01, 0.75, 0.24}
\definecolor{green-yellow}{rgb}{0.68, 1.0, 0.18}

\newcommand{\beq}{\begin{equation}}
\newcommand{\eeq}{\end{equation}}
\newcommand{\beqa}{\begin{eqnarray}}
\newcommand{\eeqa}{\end{eqnarray}}

\newcommand{\eV}{{\rm eV}}
\newcommand{\GeV}{{\rm GeV}}

\newcommand{\VPQV}{V_{\rm \cancel{\rm PQ}}}

\newcommand{\chiQCD}{\chi_{\rm QCD}}
\newcommand{\LPQV}{\mathcal{L}_{\rm \cancel{\rm PQ}}}

\newcommand{\eqn}[1]{Eq.~(\ref{#1})}

\newcommand{\N}{{\cal N}}

\begin{document}
% ----------------- preprint numbers ------------------
%\begin{frontmatter}

% ------------- Title and authors ---------------------

\title{Axion Window on New Macroscopic Forces}

\author{Luca Di Luzio}
\email{luca.diluzio@pd.infn.it}
\affiliation{\small \it Istituto Nazionale di Fisica Nucleare, Sezione di Padova, Via F.~Marzolo 8, 35131 Padova, Italy}
\author{Hector Gisbert}
\email{hector.gisbert@pd.infn.it}
\affiliation{\small \it Dipartimento di Fisica e Astronomia `G.~Galilei', Universit\`a di Padova, \\ Via F.~Marzolo 8, 35131 Padova, Italy}
\affiliation{\small \it Istituto Nazionale di Fisica Nucleare, Sezione di Padova, Via F.~Marzolo 8, 35131 Padova, Italy}
\affiliation{\small \it Escuela de Ciencias, Ingenier\'ia y Dise\~{n}o, Universidad Europea de Valencia,
Passeig de la Petxina 2, 46008 Valencia, Spain}
\author{Fabrizio Nesti}
\email{fabrizio.nesti@aquila.infn.it}
\affiliation{\small \it Dipartimento di Scienze Fisiche e Chimiche,
Universit\`a dell'Aquila, Via Vetoio, I-67100, L'Aquila, Italy}
\affiliation{\small \it Istituto Nazionale di Fisica Nucleare, Laboratori Nazionali del Gran Sasso, 
\\ I-67100 Assergi (AQ), Italy}
\author{Philip Sørensen}
\email{philip.soerensen@pd.infn.it}
\affiliation{\small \it Dipartimento di Fisica e Astronomia `G.~Galilei', Universit\`a di Padova, \\ Via F.~Marzolo 8, 35131 Padova, Italy}
\affiliation{\small \it Istituto Nazionale di Fisica Nucleare, Sezione di Padova, Via F.~Marzolo 8, 35131 Padova, Italy}

% ------------------------------------------------------
\begin{abstract}
\noindent
Axion-mediated forces are enhanced by the presence of 
CP-violating axion couplings, which are 
however
tightly constrained by electric dipole moment (EDM)
searches. We discuss the underlying hypotheses behind different sources of CP violation 
at high energies 
and the interplay between 
axion-mediated force experiments and EDM observables. 
Specifically, 
we identify various mechanisms,
based on new sources of CP violation 
or Peccei-Quinn symmetry breaking, 
that can significantly relax EDM constraints, 
leading to a substantial redefinition of the QCD axion window for axion-mediated forces. 
By considerably enlarging the QCD axion parameter space, our results provide a well-motivated target for experiments probing scalar axion couplings 
to matter fields. These include 
fifth-force tests of gravity, 
as well as
searches for spin-dependent forces 
via precision magnetometry,   
proton storage rings and  
ultracold molecules. 
\end{abstract}

\maketitle

\section{Introduction}
The Quantum Chromodynamics (QCD) axion originally emerged from the need to 
``wash out'' CP violation from strong interactions \cite{Peccei:1977hh,Peccei:1977ur,Weinberg:1977ma,Wilczek:1977pj}.
This is achieved by introducing the axion field $a$, endowed with a Peccei-Quinn (PQ) shift symmetry that renders the QCD topological angle $\theta$ unphysical.
Hence, the issue 
of CP violation in strong interactions 
is traded for 
a dynamical question about the axion vacuum expectation value (VEV), yielding an
 effective $\theta$ parameter 
\beq 
\label{eq:defthetaeff}
\theta_{\rm eff} \equiv \frac{\vev{a}}{f_a}\,,
\eeq  
where $f_a$ is the axion decay constant
and the 
electric dipole moment (EDM) of the 
neutron (nEDM)
demands
$\abs{\theta_{\rm eff}} \lesssim 10^{-10}$ \cite{Abel:2020pzs}.

A general result \cite{Vafa:1984xg} 
ensures that the QCD 
ground state energy density 
is minimal for $\theta_{\rm eff} = 0$. However, one 
relies on the fact  that QCD is a vector-like theory and does not have 
extra sources of CP violation beyond the $\theta$ term. 
Both conditions are violated in the 
Standard Model (SM),
which is a chiral theory 
that 
features a CP-violating (CPV) phase in the quark sector.  
Indeed, in Ref.~\cite{Georgi:1986kr} it was 
estimated
$\theta^{\rm SM}_{\rm eff} 
\sim 10^{-19}$
(see also \cite{Ellis:1978hq,Khriplovich:1985jr,Gerard:2012ud}).  
While this irreducible SM contribution 
to $\theta_{\rm eff}$
turns out to be too tiny to be 
testable,  
new CPV sources beyond the SM, 
generically expected to explain the baryon asymmetry of 
the universe,  
might lead to a value of 
$\theta_{\rm eff}$ closer to the nEDM bound. 

An axion VEV could also be generated from ultraviolet (UV) 
sources of PQ symmetry breaking.
This is 
motivated by the fact that the U(1)$_{\rm PQ}$, as a global symmetry, does not need to be exact and it is expected to be broken at least 
by quantum gravity effects 
(see e.g.~\cite{Kallosh:1995hi}). 

Whatever its origin, a striking consequence of 
$\theta_{\rm eff} \neq 0$ is the 
generation of a \emph{scalar} axion coupling to nucleons, 
$g^S_{aN} (\theta_{\rm eff}, \ldots) \propto \theta_{\rm eff}$ \cite{Moody:1984ba}, 
where the ellipses stand for other 
sources of CP violation,
to be discussed in this work.  
Including both scalar ($g^S_{af}$) and pseudo-scalar ($g^P_{af}$) couplings to matter fields, 
with $f = p, n, e$
(cf.~\eq{eq:Laint1}),  
one obtains different types of non-relativistic potentials 
mediated by the light axion field,  
leading to 
new macroscopic forces, 
as suggested long ago by 
Moody and Wilczek \cite{Moody:1984ba}. 
Remarkably, 
scalar axion couplings, 
being spin-independent,
strongly enhance axion-mediated forces 
compared to the case of pseudo-scalar couplings.
An updated review of axion-mediated force experiments and relevant limits can be found in 
Ref.~\cite{OHare:2020wah} (see also \cite{Raffelt:2012sp,Irastorza:2018dyq,Sikivie:2020zpn}).

While phenomenological analyses 
often focus on a 
naive definition of the QCD axion band, based on 
the simplifying assumption that $\theta_{\rm eff}$ 
is the only source of CP violation \cite{Moody:1984ba}, 
more recent studies have started to systematically 
analyze different 
sources of CP violation and their non-trivial interplay with EDM searches \cite{Pospelov:1997uv,Bertolini:2020hjc,Okawa:2021fto,Dekens:2022gha,Plakkot:2023pui,DiLuzio:2023cuk,DiLuzio:2023lmd}. 

In this work, we provide a fresh look at the UV 
origin of the $g^S_{af}$ coupling and 
identify 
different mechanisms, 
based on new sources of CP or PQ breaking, 
in order to maximize the scalar axion coupling relative to 
EDM constraints,  
thus leading 
to a substantial redefinition of the 
traditional QCD axion window 
for axion-mediated forces.  

\section{Axion couplings to matter}
Including both CP-conserving and CPV couplings,  
the axion effective Lagrangian with matter fields ($f=p,n,e$) reads%
\beq
\label{eq:Laint1}
\mathcal{L}_{af} = - g^P_{af} \, a \bar f i \gamma_5 f
- g^S_{af} \, a \bar f f
\,,
\eeq
where $g^P_{af} = C_{af} m_f / f_a$ and the dimensionless couplings $C_{af}$
are model-dependent $\mathcal{O}(1)$ numbers.   
As a benchmark scenario, we consider here the 
DFSZ model \cite{Zhitnitsky:1980tq,Dine:1981rt} 
at large $\tan\beta$, 
yielding 
$C_{ap} = 0.6$, $C_{an} = -0.3$ and 
$C_{ae} = 1/3$ 
(see e.g.~\cite{DiLuzio:2020wdo} for a derivation). 
The QCD axion mass and decay constant 
follow the standard relation 
$m_a \simeq 5.7 \, ( 10^{9}\,\GeV / f_a)
\, \eV$. 

In the traditional QCD axion scenario with a nonzero $\theta_{\rm eff}$, for $g_{aN}^S$ $(N=p,n)$ one can use 
the standard isospin-symmetric  formula of Moody and Wilczek~\cite{Moody:1984ba},  
with the correct extra 1/2 factor~\cite{Bertolini:2020hjc}, which gives at most
\bea
\label{eq:gstheta}
g_{aN}^{S,\theta}&\simeq&
2\times 10^{-21} \left(\frac{10^9 \,\GeV}{f_a}\right) ,
\eea
assuming maximal $\theta_{\text{eff}}=1.2\times 10^{-10}$, saturating the $d_n$ constraint. 
The value \eqref{eq:gstheta} defines the upper 
side of the traditional QCD axion band for 
axion-mediated forces, currently 
employed for instance in \cite{AxionLimits}. The lower side of the band 
is estimated through the irreducible SM contribution, via the CKM phase, as \cite{Okawa:2021fto}
\beq 
\label{eq:gsthetaCKM}
g_{aN}^{S,\text{CKM}} \simeq 10^{-30} \left(\frac{10^9 \,\GeV}{f_a} \right) . 
\eeq
Note, however, that a CPV scalar coupling to nucleons $g^S_{aN}$ 
can arise from various mechanisms related to new
sources of CP or PQ violation. These can lead to $a)$ new terms in the meson plus axion chiral Lagrangian ($\chi$PT), that induce a shift of the axion and chiral vacuum leading to  CPV couplings~\cite{Pich:1991fq}, but also to $b)$ direct CPV axion-nucleon terms in the baryon chiral Lagrangian (B$\chi$PT).

Experimental searches typically operate at the atomic level rather than directly probing $g_{af}^S$.  When considering an atomic system  $^{A}_{Z}X$, 
the 
axion scalar coupling 
is effectively described by the 
 average 
\bea\label{eq:gaN}
g^S_{aX}
\equiv\frac{A-Z}{A}\,g_{an}^S\,+\frac{Z}{A}\,g_{ap}^S\,+\frac{Z}{A}\,g_{ae}^S\,.
\eea
In our theoretical predictions, we will consider 
the case of Tungsten $(A = 184, Z = 74)$ 
which pertains to ARIADNE \cite{Arvanitaki:2014dfa,ARIADNE:2017tdd}. 
Other elements,  
relevant for fifth-force searches, 
imply a relative variation 
from the Tungsten benchmark 
that is below the $10\%$ level.

In the following, we explore the possibility of maximizing this scalar axion-matter coupling in comparison to  $g_{aX}^{S,\theta}$, equal to the traditional $g_{aN}^{S,\theta}$ in the isospin limit,  while  satisfying the EDM constraints.

\section{CPV  couplings from $\boldsymbol{\chi}$PT}
In the $\chi$PT Lagrangian, the  terms generated by CP or PQ breaking lead to a realignment of the vacuum, namely a nonzero
$\theta_{\rm eff}$ together with  
$\pi_0$, $\eta_8$ and $\eta_0$
meson VEVs if chiral symmetry is also broken. These four vacuum shifts induce CPV baryon couplings in the B$\chi$PT Lagrangian, 
reported for completeness in \app{app:BchiPT}. 
It is important to note that the induced couplings generally depend on just \emph{three} combinations.  Indeed, the B$\chi$PT Lagrangian feels the vacuum realignment only via the three quark masses, the only spurions of chiral $SU(3)_V\times U(3)_A$. 
As a result, all effects 
depend on the 
following three chiral phases (with $q=u,d,s$):
\bea
\label{eq:alpha}
\!\!\!\!\!\alpha_{q}\!&=&\!\!\left[\frac{\lambda_3}2 \frac{\left\langle \pi   _0\right\rangle}{F_\pi}+\!\frac{\lambda_8}2\frac{\langle \eta_8 \rangle}{F_\pi} +\!\frac{\mathbf 1}{\sqrt{6}} \frac{\left\langle \eta _0\right\rangle}{F_\pi}\right]_{qq}
\!\!+\frac{m_*\theta_{\text{eff}} }{2\,
   m_q},\ 
\eea
where the $\lambda$'s are the Gell-Mann matrices, $m_*=(m_u^{-1}+m_d^{-1}+m_s^{-1})^{-1}$, and the pion  decay constant is $ F_\pi \simeq 92$\,MeV.
For the neutron and proton CPV couplings, one can 
rewrite the results of~\cite{Bertolini:2020hjc} (see also~\cite{Cirigliano:2016yhc,Bertolini:2019out}) in a simple and general way in terms of $\alpha$'s: 
\begin{align}
\label{eq:gaNS}
g^S_{ap,\,n}\! &=\!-\frac{8 m_*B_0  }{f_a}\bigg(b_0{\normalsize \sum_q} \alpha_q
   +b_+
    \alpha_{u,d}
   +b_-
    \alpha _s 
    \bigg) \\ 
&\simeq 10^{-11} \frac{10^9\,\GeV}{f_a}(2.4\alpha_{u,d} + 2.8 \alpha_{d,u}+1.5 \alpha_s ) 
    \,, \nonumber
\end{align}
with 20\% hadronic uncertainties, 
see \app{app:BchiPT} for details.
Expressions in terms of $\alpha$'s can be derived also for 
nucleon-meson CPV couplings (cf.\ Appendix~\ref{app:BchiPT})
so that this fact holds in general 
even for EDM observables. 
We thus find that from $\chi$PT there is a model-independent correlation between  EDMs and $g^S_{ap,n}$, all driven by  three phases $\alpha_{u,d,s}$.

\section{EDM constraints on CPV couplings}
We focus first on the CPV effects induced in $\chi $PT, considering 
the $g_{an,p}^S$ terms in \eqn{eq:gaN}, with $g_{ae}^S$ set to zero.
The strongest EDM limits are currently imposed by  neutron and Mercury.
For them, we find the constraints
\begin{align}
\label{eq:dn}
\frac{d_n}{d_n^<} &\,\simeq\, 10^{10} 
| 3.7\, \alpha_u+5.9\, \alpha_d+40.0\, \alpha_s|  \lesssim 1 \,,\\
\label{eq:dhg}
\frac{d_{\text{Hg}}}{d_{\text{Hg}}^<} &\,\simeq\, 10^{10} |3.1\, \alpha_u+5.0\, \alpha_d+36.0\, \alpha_s| \lesssim 1\,,\quad
\end{align}
with 10\%  and 50\% hadronic uncertainties respectively, 
see \app{app:BchiPT} for details.
Here, $d_n^<=1.8 \times 10^{-26}\,e\,\text{cm}$~\cite{Abel:2020pzs} and $ d_{\text{Hg}}^<=6.3 \times 10^{-30}\,e\,\text{cm}$~\cite{Graner:2016ses} are the 
experimental bounds on the
neutron and Mercury EDM, respectively.

While other nuclei lead to weaker bounds,
more constraints arise from EDMs of paramagnetic systems from the contribution of two-photon exchange processes between electrons and the nucleus induced by CP-odd semileptonic interactions~\cite{Flambaum:2019ejc}. 
The most stringent one, set by paramagnetic ThO~\cite{ACME:2018yjb},
leads to the additional constraint (cf.~\app{app:BchiPT})
\bea
\label{eq:dtho}
10^{10}|
0.11\, \alpha_u-0.22\, \alpha_d|  \lesssim 1 \, ,
\eea
with 10\% hadronic uncertainties, 
see \app{app:BchiPT} for details.
The limit in Eq.~\eqref{eq:dtho} is slightly weaker, but together with Eqs.~\eqref{eq:dn} and \eqref{eq:dhg}, constrain all $\alpha$'s to be of the order of $10^{-9\div 10}$. 
This implies that the axion scalar coupling 
in \eq{eq:gaNS}
cannot deviate much from the traditional value. 
Including uncertainties, 
we find at most 
$|g^S_{aX}/g^{S,\theta}_{aX}|\lesssim \mathcal{O}(10)$.

In conclusion, the three EDM constraints fix the three phases, impeding a relevant enhancement of $g_{aX}^S$ relative to the traditional QCD axion case $g_{aX}^{S,\theta}$. This result is model-independent, as far as  $\chi$PT-induced effects are concerned.

The situation could potentially change with 
direct contributions to the B$\chi$PT Lagrangian. 
New terms in general lead to additional combinations different 
from those in Eq.~\eqref{eq:alpha}, possibly 
evading the EDM limits.   
The possible new operators,  classified in~\cite{Dekens:2022gha},  suffer from uncertainties in the  low-energy constants (LECs), not determined by chiral symmetry, and a 
model-independent analysis is premature. 
One can, of course, imagine that 
more operators conspire to 
bypass the EDM constraints while allowing for a larger $g_{aX}^S$.
This
scenario, hard to obtain in predictive theories, could be checked in specific models.

To this aim, theoretical work on the LECs would be most welcome,
while on the experimental side future improvements on EDMs such as Radon~\cite{Bishof:2016uqx} will also be important to sharpen the connection between EDMs and $g_{aN}^S$. 

\section{Semi-leptonic operators}
We turn here to CPV effects induced only by the scalar axion-electron coupling $g_{ae}^S$, the third term in Eq.~\eqref{eq:gaN}. This coupling can be generated by semileptonic operators as
\begin{equation} 
\mathcal{L}_\text{sl}=C^P_{eq}\,(\bar e_L e_R)(\bar q \gamma_5 q)
+\text{h.c.} \,,
\label{eq:semiL}
\end{equation}
with implicit sum over $q=u,d,s$.  
After the canonical axion-dependent chiral rotation of the quarks (cf.~\app{app:BchiPT}), 
\eq{eq:semiL} leads to~\cite{Dekens:2022gha}
\begin{equation} \label{eq:semiN}
g_{ae}^S=\frac{m_*\,B_0\,F_\pi^2 }{2\,f_a}\,{\rm Im}\sum_q\frac{1}{m_q} C^P_{eq}\,.
\end{equation}
The interactions in \eq{eq:semiL} also generate 
4-fermion operators involving nucleons and electrons \cite{Dekens:2022gha}
\bea  
\label{eq:LeN}
\!\!\!\!\!\!\!\!\!\!\!\! \mathcal{L}_{eN} & =&-\frac{G_F}{\sqrt{2}}\bigg\{i\, \bar e \g_5 e\, \bar N\left(C_S^{(0)}+\tau_3 C_S^{(1)}\right) N \nonumber \\
&+& \bar e e\, \frac{\partial_\mu}{m_N} \left[\bar N\left(C_P^{(0)}+\tau_3 C_P^{(1)}\right)S^\mu N\right]
\bigg\} \,,
\eea
where $N= (p\,\,n)^T$   is the nucleon doublet, $S^\mu$ is its spin, $G_F$ is the Fermi constant,   and $C_{S,P}^{(0,1)}$ are the Wilson coefficients encoding the short-distance coefficients 
$C^{P}_{eq}$ of Eq.~\eqref{eq:semiL}. The Mercury EDM or CPV effects induced in polar molecules like ThO~\cite{ACME:2018yjb}, YbF~\cite{Hudson:2011zz}, and HfF~\cite{Cairncross:2017fip} receive contributions from  $C_{S,P}^{(0,1)}$. Using Eq.~(25) in Ref.~\cite{Dekens:2022gha}, we find the conditions on the $C^P_{eq}$ to cancel them:
\begin{equation}
C^P_{eu}=C^P_{ed}=C^P_{es}\equiv C\,, 
\label{eq:cond}
\end{equation}
for some $C$.
These conditions still 
allow
for a non-zero $ g_{ae}^{S}$ from \eq{eq:semiN}:
\begin{align}
    g_{ae}^{S} &=-\frac{B_0\,F_\pi^2 }{2\,f_a}{\rm Im}\,C
    \,.\label{eq:gaetunhad}
\end{align}
Hence, strikingly, the EDM limits are evaded for flavour universal semileptonic couplings.
The coefficient $\text{Im}\,C$ 
is
constrained by 
the high-$p_T$ tails of the 
$pp \to \ell\ell$ Drell-Yan processes. The results in Ref.~\cite{Allwicher:2022gkm} imply 
$\text{Im}\,C \lesssim (2 \, \text{TeV})^{-2}$, 
which allows for a substantial enhancement of the scalar axion coupling, 
$|g_{aX}^{S}/g_{aX}^{S,\theta}|\lesssim 1\times 10^3$.

\section{PQ-breaking as the origin of CPV couplings}
Another possibility to generate a 
scalar axion coupling to fermions is to consider a source 
of PQ breaking, which also breaks CP in the axion sector. 
A common choice is provided by operators of the type
$\LPQV \supset - e^{i\delta} \phi^n \Lambda^{-n+4}$, 
with $\delta$ a generic phase,    
$\Lambda$ is the UV scale at which the operator is generated, 
and $\phi=\frac{f_a}{\sqrt{2}} e^{ia/f_a} $ is a complex scalar where the axion is the 
angular mode and the radial mode is integrated out. 
Including the QCD axion potential, 
$V_{\rm QCD} = - \chi_{\rm QCD} \cos\frac{a}{f_a}$ 
with $\chi_{\rm QCD} \approx (76\, \text{MeV})^4$, 
the induced axion VEV reads 
$\theta_{\rm eff} 
	\simeq - 2^{1-\frac{n}{2}} n \Lambda^{4-n} f_a^n \sin \delta / \chiQCD$. 
Note that such a PQ-breaking scenario directly matches 
to the traditional QCD axion coupling in \eq{eq:gstheta}.

Another class of operators, recently discussed in 
Refs.~\cite{Zhang:2022ykd,Zhang:2023gfu,DiLuzio:2024fyt},  
takes the generic form $(\phi/\Lambda)^n \mathcal{O}_{\rm SM}$, where $\mathcal{O}_{\rm SM}$ is an operator made of SM fields. 
To maximize the contribution to 
$g^S_{aX}$ in \eq{eq:gaN}, relative to EDM bounds, we
consider a $\phi^n$ coupling to the electron Yukawa,\footnote{Similar 
operators 
with $\phi^n$ coupled to light quarks or gluons 
yield an unsuppressed long-distance contribution to $\theta_{\rm eff}$ \cite{Zhang:2022ykd}, 
 preventing a significant   
contribution to $g^S_{aX}$.} 
\begin{equation}
\LPQV
\supset - e^{i\delta} \left(\frac{\phi}{\Lambda}\right)^{\!\!n}\!\! \frac{\sqrt{2}m_e}{v} \bar{L}_{L} H e_R+\text{h.c.} \, , 
\label{eq:fermion Yukawa}
\end{equation}
where $v=246$ GeV. This operator generates a 
scalar axion-electron coupling directly at tree level
\begin{gather}
    g^S_{ae} = n\left(\frac{f_a}{\sqrt{2}\Lambda}\right)^n \frac{m_e}{f_a} \sin\delta \, , \label{eq:gSae}
\end{gather}
and, 
after setting $\vev{H} = ( 0 \ \ v/\sqrt{2})^T$, 
it also leads to an 
axion tadpole, $\VPQV = -\sigma a + \ldots $, 
at one loop. 
In the leading-log approximation, we find 
\beq 
\sigma (\mu) = \frac{n}{2\pi^2} 
\left(\frac{f_a}{\sqrt{2}\Lambda}\right)^{\!n} 
\frac{m_e^4}{f_a} \sin\delta \ln \( \frac{v}{\mu} \) ,
\eeq
where $\mu$ denotes the renormalization scale. 
Including the 
contribution of the QCD axion potential, 
we obtain the induced axion VEV
\begin{align}
\label{eq:thetaeffindYuke}
\theta_{\rm eff} &\simeq  
 \frac{n}{2\pi^2} 
 \left(\frac{f_a}{\sqrt{2}\Lambda}\right)^{\!n} 
 \!\frac{m_e^4}{\chiQCD}
 \sin\delta 
 \,\ln\left(\frac{v}{1 \, \text{GeV}}\right) 
 \, ,  
\end{align}
where we took $\mu = 1$ GeV 
to assess the 
nEDM bound. 
By using \eq{eq:thetaeffindYuke} 
to express \eq{eq:gSae} 
in terms of $\theta_{\rm eff}$, we obtain
\beq
\label{eq:ratiogtheta}
g^S_{ae} = 
\frac{2\pi^2}{\ln\(\frac{v}{1 \, \text{GeV}}\)} 
\frac{\chiQCD}{m_e^3 f_a}  \theta_{\rm eff} 
\, . 
\eeq
The nEDM bound, $|\theta_{\rm eff}| \lesssim 10^{-10}$, provides the main limiting 
factor for this coupling,\footnote{Other direct contributions to EDMs come from  
axion exchange involving $aG\tilde G$ and 
$a \overline{e} e$ vertices, leading to semileptonic 
operators as in \eq{eq:LeN}. Using results from 
\cite{DiLuzio:2020oah},  
we obtain $g^S_{ae} \lesssim 2.2 \, (f_a / 10^9 \, \text{GeV})$ 
from ThO EDM, which is negligible compared to the bound from 
nEDM in \eq{eq:ratiogtheta}.}
which in this scenario dominates by far the
atomic average in \eq{eq:gaN}, allowing for $|g_{aX}^{S}/g_{aX}^{S,\theta}|\lesssim 2\times 10^7$.

\section{$Z_{\N}$ axion} 
The last possibility 
that we consider 
is a modification of the standard $m_a$--$f_a$ QCD relation, suppressing $m_a$ 
for fixed $f_a$. 
This can be achieved by employing 
$\N$ mirror copies of the SM \cite{Hook:2018jle,DiLuzio:2021pxd,DiLuzio:2021gos},
with $\N$ odd and $\text{SM}_k \to \text{SM}_{k+1(\text{mod}\, \N)}$
under the $Z_{\N}$ symmetry 
and the axion 
acting non-linearly: 
$a \to a + 2\pi k / \N$, with $k = 0,\ldots, \N-1$. 
In this way, the axion potential gets exponentially suppressed 
and, in the large $\N$ limit,  
the axion mass 
relative to the standard QCD axion mass 
scales as \cite{DiLuzio:2021pxd}
\beq 
\frac{(m_a)_{\N}^2}{m_a^2} 
\simeq \frac{\sqrt{1-z^2}(1+z)}{\sqrt{\pi}} \N^{3/2} 
z^{\N -1} \, ,
\eeq
where $z = m_u / m_d \simeq 0.48$. 
In these scenarios, 
an important constraint on $f_a$ arises 
from the fact that the exponential axion mass 
suppression is spoiled by finite density effects 
in stellar environments \cite{Hook:2017psm}.  
In particular, 
the strongest constraints arise from the 
modifications of the mass-radius relationship of white dwarfs, which exclude $33 \leq \N \leq 69$ \cite{Balkin:2022qer}. 
Assuming $\N\leq 31$, 
we find 
$|g_{aX}^{S}/g_{aX}^{S,\theta}|\lesssim 5\times 10^3$.

\section{Axion-mediated force experiments}
\begin{figure}[t!]
    \centering
   \includegraphics[scale=0.20]{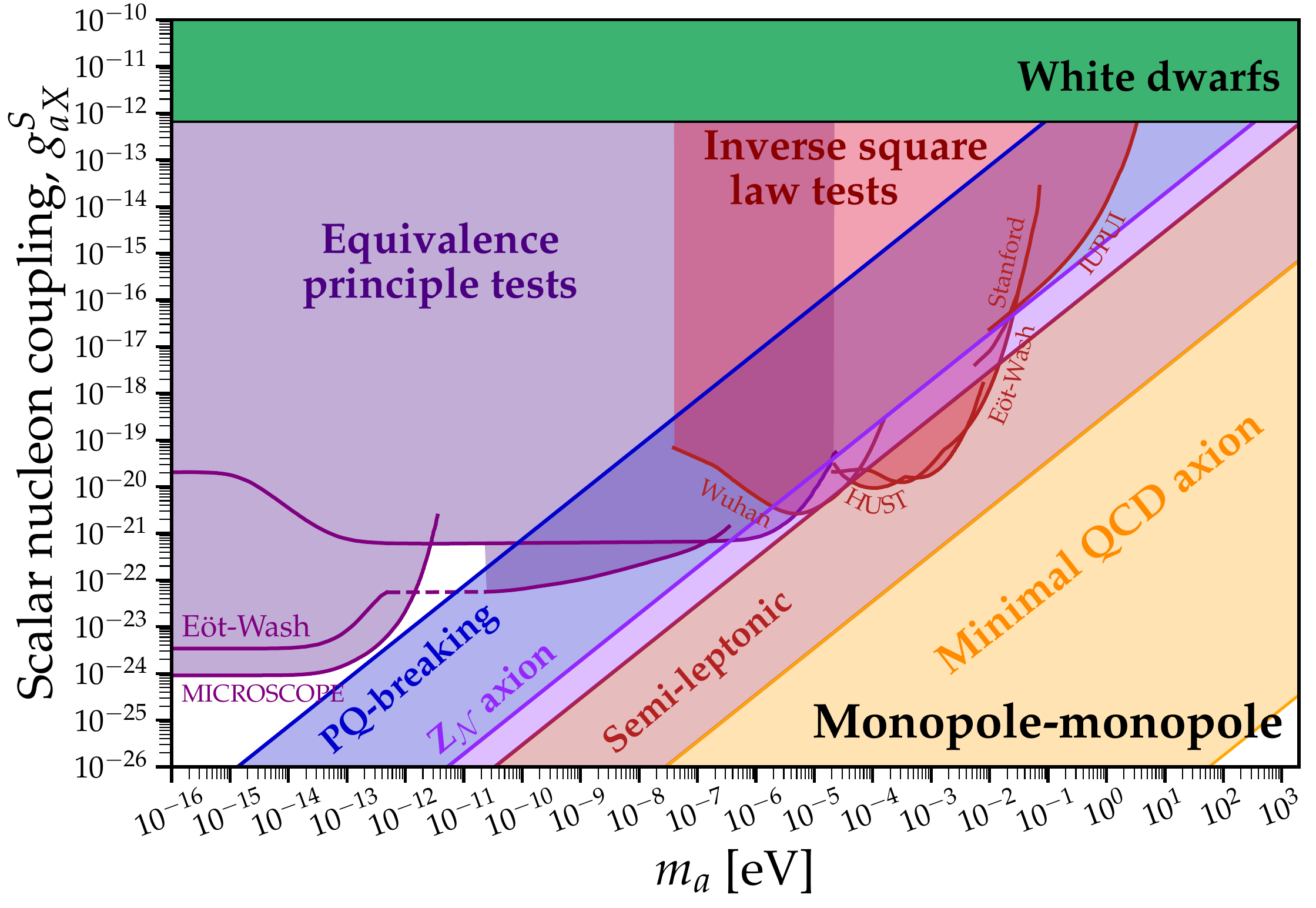}
   \includegraphics[scale=0.20]{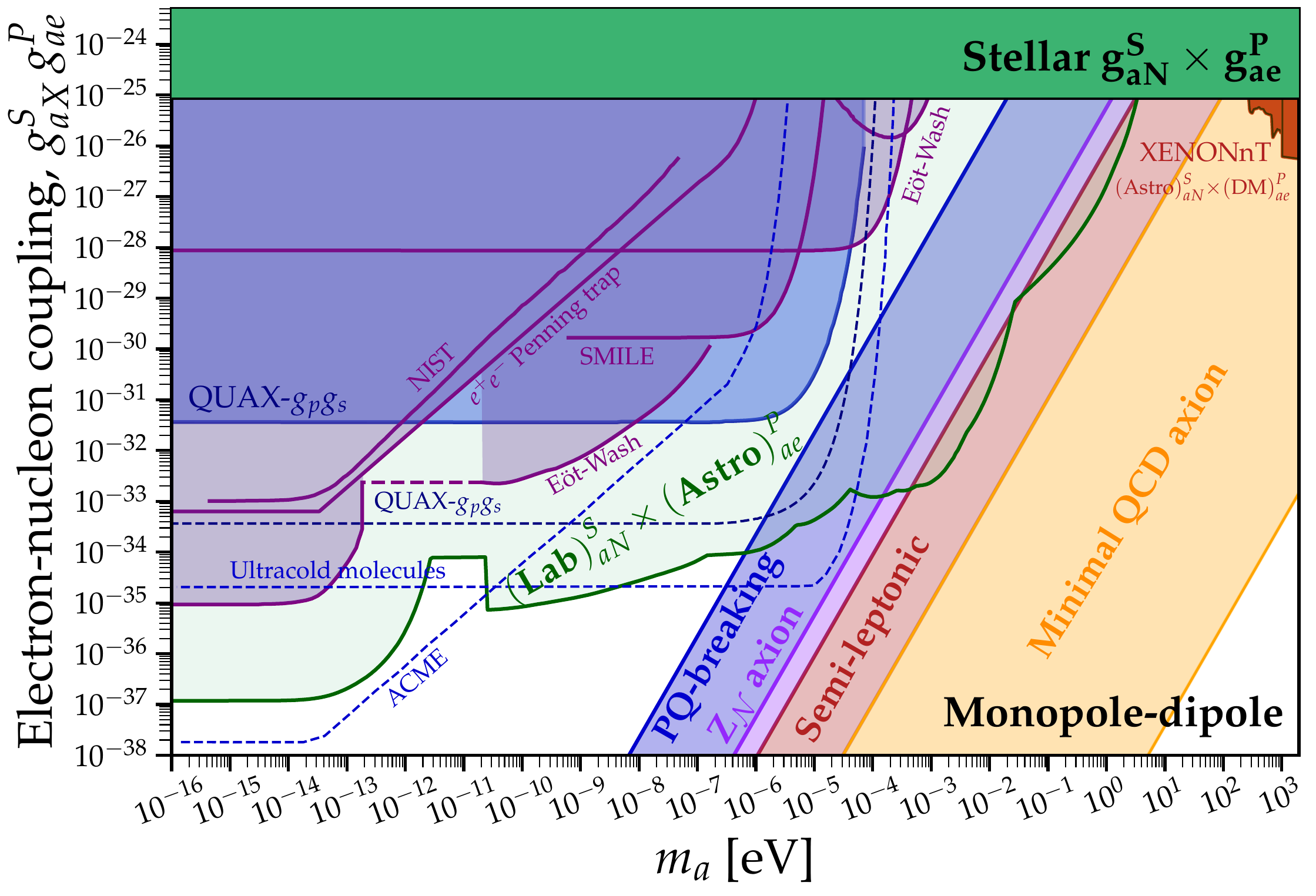}
   \includegraphics[scale=0.20]{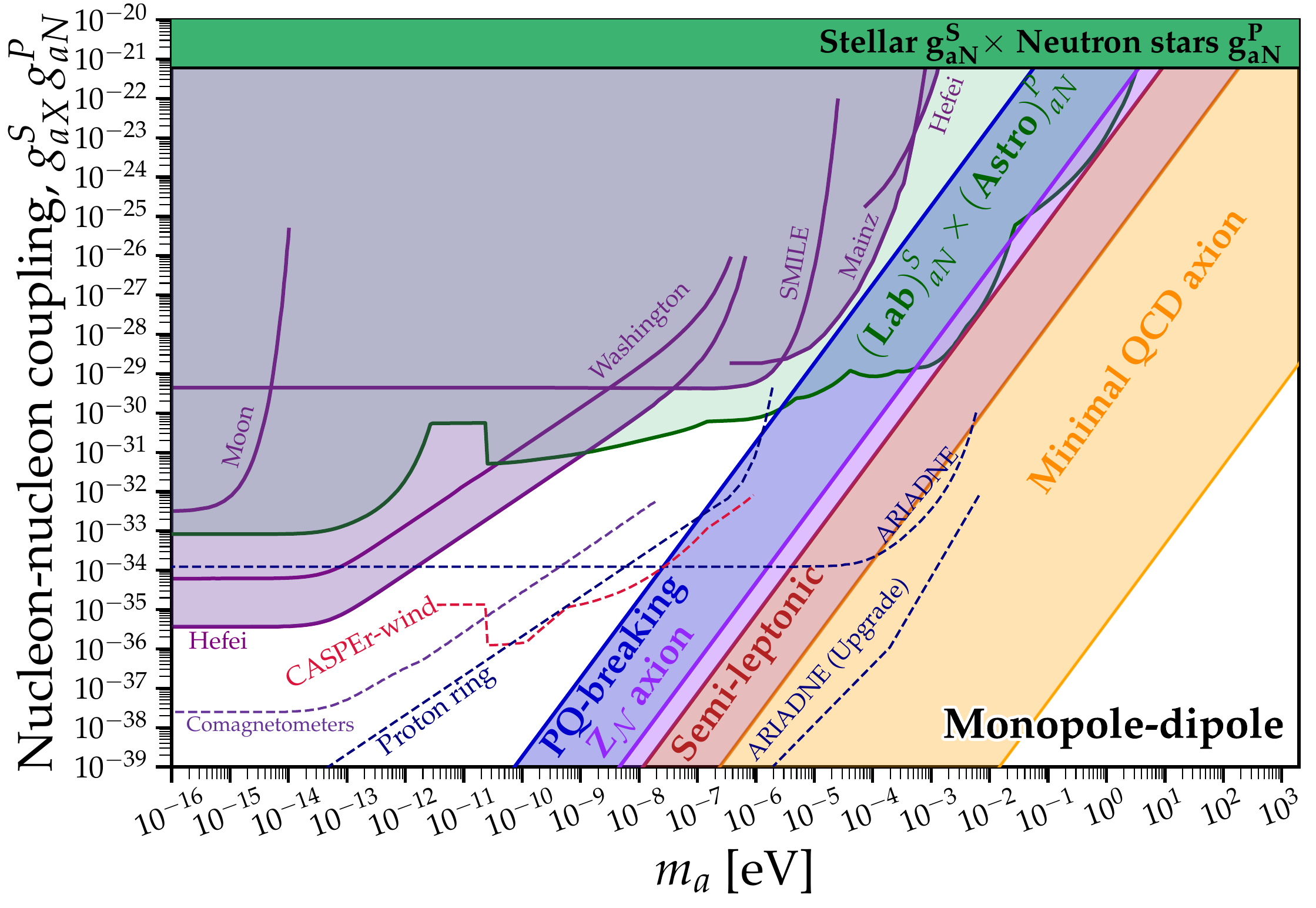}
   \caption{QCD axion window
   for axion-mediated forces vs.~current constraints (full lines) and future experimental sensitivities (dashed lines). 
   Here $g^S_{aX}$ is the effective scalar axion-atom coupling defined in Eq.~\eqref{eq:gaN}.
   Figure adapted from \cite{AxionLimits,OHare:2020wah}. 
   }
   \label{fig:interplay}
\end{figure}

Depending on the combination of couplings involved, 
axion-mediated non-relativistic 
potentials can be of three types: 
$g^S_{af} g^S_{af}$ (monopole-monopole),
$g^S_{af} g^P_{af}$ (monopole-dipole),
or $g^P_{af} g^P_{af}$ (dipole-dipole). 
The idea of searching for dipole-dipole axion interactions   
in atomic physics 
is as old as the axion itself \cite{Weinberg:1977ma}.
However, dipole-dipole forces turn out to be 
spin-suppressed 
and suffer from large backgrounds from ordinary magnetic forces. Furthermore, our models enhance only 
scalar couplings and thus do not improve the situation for dipole-dipole interactions. Hence, in the following, we will focus 
on monopole-monopole and monopole-dipole interactions, 
whose parameter space is displayed in Fig.~\ref{fig:interplay}. 

In this figure, the conventional QCD axion region is 
represented by an orange band, whose upper (lower) side is set by the scalar coupling in \eq{eq:gstheta} (\eq{eq:gsthetaCKM}), 
while for the pseudo-scalar couplings we employed 
the predictions of the DFSZ model at large $\tan\beta$ (cf.~discussion below \eq{eq:Laint1}). 
Among the models discussed in this work, 
we consider three examples, given respectively by  
the inclusion of CPV
semi-leptonic operators (red), the $Z_\mathcal{N}$ axion model with modified $m_a$--$f_a$ relation (purple), 
and the PQ-breaking electron 
Yukawa scenario (blue).   
These extend the conventionally expected range of scalar couplings. The figure should be understood so that the bands overlap from below, such that e.g.~the PQ-breaking model predicts a coupling between the upper limit of the blue band and down to the lower limit of the minimal QCD axion band (orange). 
See previous sections for how the upper limits on the 
scalar coupling is obtained in non-minimal QCD axion models.

The top panel of Fig.~\ref{fig:interplay} 
displays searches based on monopole-monopole interactions.
In contrast with gravity, these long-range forces deviate from the inverse-square law and depend on the material's composition. 
Tests of the inverse-square law
\cite{Chen:2014oda,Kapner:2006si,Lee:2020zjt,Geraci:2008hb,Yang:2012zzb,Tan:2020vpf,Ke:2021jtj} 
can reach parameter space from all three models presented here.
At longer ranges, i.e.~lower axion masses, composition dependence becomes more important than deviations from the inverse-square law. This enables searches for violations of the equivalence principle \cite{Smith:1999cr,Berge:2017ovy} to probe the parameter space offered by the $Z_{\N}$ and the PQ-breaking models.

These results 
suggest that improvements in tests of the equivalence principle or the inverse square law have the chance to discover 
these particular types of QCD axions.

Another interesting opportunity is given by monopole-dipole searches, whose limits (full lines) 
and projected sensitivities (dashed lines) 
are reported in the 
middle and lower panels of Fig.~\ref{fig:interplay} for the case of pseudo-scalar axion 
couplings to electrons ($g_{ae}^P$) and nucleons ($g_{aN}^P$), respectively. 
In particular, ARIADNE \cite{Arvanitaki:2014dfa,ARIADNE:2017tdd}
aims to probe the monopole-dipole force 
generated by a rotating source mass 
and detected via nucleon spins, 
employing 
precision magnetometry. 
A similar approach is pursued by QUAX-$g_pg_s$
\cite{Crescini:2016lwj,Crescini:2017uxs,Crescini:2020ykp}, 
using instead electron spins. Several other experiments have established bounds or are attempting to improve the sensitivity to either 
$g_{aX}^S g_{ae}^P$
\cite{Wineland:1991zz,Heckel:2008hw,Hoedl:2011zz,Terrano:2015sna,Stadnik:2017hpa,Lee:2018vaq,Dzuba:2018anu,Fan:2023hci,Agrawal:2023lmw,Baruch:2024frj,Baruch:2024fbh} or 
$g_{aX}^S g_{aN}^P$
\cite{Wei:2022ggs,Agrawal:2022wjm,Baruch:2024frj,Baruch:2024fbh} couplings.
Regardless of which pseudo-scalar coupling is employed, our scenarios considerably 
improve the prospects for monopole-dipole searches. The projected reaches of QUAX-$g_p g_s$ as well as proposals based on ultracold molecules \cite{Agrawal:2023lmw} and proton storage rings \cite{Agrawal:2022wjm} can probe parameter space beyond the traditional QCD
band which is populated by the models studied here, 
while
ARIADNE
would be able to test wide ranges of parameter space, 
also inside the minimal QCD axion band.\\

\section{Conclusions}
In this paper, we have reconsidered the UV origin of the 
$g^S_{af}$ coupling and identified various mechanisms based on new sources of CP or PQ breaking. These mechanisms can significantly relax EDM constraints, leading to a substantial redefinition of the traditional QCD axion window for axion-mediated forces, 
as shown in Fig.~\ref{fig:interplay}. 
Our findings suggest an expanded QCD axion parameter space, providing strong motivation for pushing the current limits on 
the search for new macroscopic forces. 

\begin{acknowledgments}
We thank Antonio Rodríguez-Sánchez and Alessio Maiezza for useful discussions. The work of LDL, HG and PS is supported
by the European Union -- NextGeneration 
EU and by the University of Padua under the 2021 
STARS Grants@Unipd programme 
(Acronym and title of the project: CPV-Axion -- Discovering the CP-violating axion)
as well as by the 
European Union -- Next Generation EU and
by the Italian Ministry of University and Research (MUR) 
via the PRIN 2022 project n.~2022K4B58X -- AxionOrigins. 
\end{acknowledgments}

\onecolumngrid

\clearpage
\appendix

\section{Baryon-meson couplings contributions to EDMs}
\label{app:BchiPT}

In this Appendix, we discuss the contributions of baryon-meson couplings to the neutron and proton EDMs. 
The leading order (LO) baryon chiral Lagrangian can be
written as
\begin{align}    
\mathcal{L}_{\rm B}&={\rm Tr}\left[\bar{B}i\gamma^\mu(\partial_\mu B+[\Gamma_\mu,\,B])-M_B \bar{B}B\right]
-\frac{D}{2}{\rm tr}\left[\bar{B}\gamma^\mu\gamma_5\{\xi_\mu,\,B\}\right]
-\frac{F}{2}{\rm tr}\left[\bar{B}\gamma^\mu\gamma_5[\xi_\mu,\,B]\right]
\nonumber \\[1.0ex]
&-\frac{\lambda}{2} \ {\rm tr}\left[\xi_\mu\right]{\rm tr}\left[\bar{B}\gamma^\mu\gamma_5B\right]+b_D \ {\rm tr}\left[\bar{B}\{\chi_+,\,B\}\right]
+b_F \ {\rm tr}\left[\bar{B}[\chi_+,\,B]\right]+b_0 \ {\rm tr}\left[\chi_+\right]{\rm tr}\left[\bar{B}B\right] \,,
\label{Lbaryon}
\end{align}
where 
we employed the definitions 
\begin{align}
&B=\begin{pmatrix}
      \frac{1}{\sqrt{2}}\Sigma^0+\frac{1}{\sqrt{6}}\Lambda^0 & \Sigma^+ & p \\
      \Sigma^- & -\frac{1}{\sqrt{2}}\Sigma^0+\frac{1}{\sqrt{6}}\Lambda^0 & n \\
      \Xi^- & \Xi^0 & -\frac{2}{\sqrt{6}}\Lambda^0 \\
   \end{pmatrix},\hspace{2em}
U=\xi_R\,\xi_L^\dagger    \qquad
(\xi_R=\xi_L^\dagger)~,
\end{align}
and
%\begin{align}
$ \Gamma_\mu\equiv \frac{1}{2}\xi_R^\dagger(\partial_\mu-i\,r_\mu)\xi_R+\frac{1}{2}\xi_L^\dagger(\partial_\mu-i\,l_\mu)\xi_L$,
% \\[1.5ex]
$ \xi_\mu\equiv  i\xi_R^\dagger(\partial_\mu-i\,r_\mu)\xi_R-i\xi_L^\dagger(\partial_\mu-i\,l_\mu)\xi_L$,
%\\[2ex]
$ \chi_+\equiv  \xi_L^\dagger \chi \xi_R+ \xi_R^\dagger \chi^\dagger \xi_L$.
%\end{align}

Here, $M_B$ denotes the baryon mass in the chiral limit, while $\chi=2 B_0 \ {\rm diag}\{m_u,m_d,m_s\}$ is the chiral spurion including the quark masses.
In \eq{Lbaryon}, the interaction terms
proportional to $D$, $F$ and $\lambda$ are CP conserving, while those proportional to $b_D$, $b_F$
and $b_0$ violate CP. The   $D$ and $F$ LECs are extracted from baryon semi-leptonic decays, and they read at tree level $D \simeq 0.8$ and $F \simeq 0.5$~\cite{Scherer:2012xha}. The LEC $B_0$ is given by $B_0=m_\pi^2/(m_d+m_u)$, while $b_{D,F}$ 
are determined from the baryon octet mass splittings,
$b_D\simeq 0.07\,\rm GeV^{-1}$, $b_F\simeq -0.21\,\rm GeV^{-1}$ at LO~\cite{Pich:1991fq}.  The value
of $b_0$ is determined from the pion-nucleon sigma-term as $b_0\simeq -\sigma_{\pi N}/4m_\pi^2$.
From~\cite{Hoferichter:2015dsa,Hoferichter:2016ocj} one obtains
$b_0= -0.76\pm 0.04\, \rm GeV^{-1}$ at 90\% confidence level. The LEC $\lambda$ is unknown and is set to zero. We also defined the shorthands $b_\pm=b_D\pm b_F$.

\smallskip

The QCD $\theta$-term plus the axion is 
rotated away by an appropriate axion-dependent chiral rotation of the quark fields,
$
q_L\to e^{- i\,\a_q}q_L$, $
q_R\to e^{ i\,\a_q}q_R$,
 with
$\a_q=\left(\theta+\frac{a}{f_a}\right)m_*/2m_q$~\cite{Pich:1991fq}.
By applying this diagonal $U(3)_A$ field transformation on the $\xi_{L,R}$ and the baryon chiral fields,  the axion
 is effectively included in the meson Lagrangian as well as in the baryon Lagrangian \eqref{Lbaryon}. With this  choice of $\alpha_q$ the axion does not mix with
$\pi^0$ and $\eta_8$.  

\smallskip

In the presence of new physics in the $\chi$PT Lagrangian, one has to further rotate the $U$ fields to have $\vev{U}=1$, as well as to shift the axion field. This again amounts to a diagonal rotation (we consider strangeness preserving interactions here) so that one can consider altogether three generic $\alpha_q$,  incorporating the mesons and axion VEVs, as in \eq{eq:alpha} in the main text.  As 
already mentioned,
this \emph{global} chiral rotation is effectively equivalent to 
a rotation of the 
spurions of chiral $SU(3)_V\times U(3)_A$, 
namely a rotation of the quark masses in the meson and baryon chiral Lagrangian \eqref{Lbaryon}, as follows: $m_q\to m_q e^{2i\alpha_q}$. 

It is this phase rotation that generates, from the baryon Lagrangian, the various CPV baryon interactions with the physical meson fields~\cite{Pich:1991fq,An:2009zh}.
We parametrize their couplings as 
$\mathcal{L}_{\text{B}\chi\text{PT}}^{\rm \,CPV}\,\supset\,\bar{g}_{ABC}\,A\,B\,C$, where $A$, $B$  are baryon fields and $C$ is a meson, and their expressions are reported below.

The neutron and proton EDMs are generated at loop level and can be expressed, following e.g.\ Ref.~\cite{Guo:2012vf},  in terms of these couplings and loop functions, which are expanded at leading order in the meson masses.  We write
$d_{n,p}\,=\,\sum_i\,d_{n,p}^{(i)}$, where the index $i$ runs over all baryon-meson couplings contributing to the neutron and proton EDMs. Specifically, for the neutron EDM, only two couplings contribute~\cite{Guo:2012vf}:
\begin{align}
    d_n^{(\bar{n}p\pi^-)}\,&=\,\frac{e\,(F\,+\,D)\,\bar{g}_{\bar{n}p\pi^-}\,(\pi \,m_\pi\,-\,2\,m_N\,\log(m_\pi^2/m_N^2))}{16\,\sqrt{2}\,F_\pi\,m_N\,\pi^2}~,\\
    d_n^{(\bar{n}\Sigma^- K^+)}\,&=\,\frac{e\,(F\,-\,D)\,\bar{g}_{\bar{n}\Sigma^- K^+}\,(\pi\,m_K\,-\,2\,m_N\,\log(m_K^2/m_N^2))}{16\,\sqrt{2}\,F_\pi\,m_N\,\pi^2}~,
\end{align}
while more couplings contribute to the proton EDM due to its electric charge:
\begin{align}
    d_p^{(\bar{p}p\pi^0)}\,&=\,-\,\frac{e\,(D\,+\,F)\,\bar{g}_{\bar{p}p\pi^0}\,m_\pi}{16\,\sqrt{2}\,F_\pi\,m_N\,\pi}~,&    d_p^{(\Sigma^+\bar{p}K^0)}\,&=-\frac{e\,(D\,-\,F)\,\bar{g}_{\Sigma^+\bar{p}K^0}\,m_K}{8\,\sqrt{2}\,F_\pi\,m_N\,\pi}~,\\  
    d_p^{(\bar{p}p\eta_8)}\,&=\,-\,\frac{e\,(D\,-\,3\,F)\,\bar{g}_{ \bar{p}p\eta_8}\,m_{\eta_8}}{48\,\sqrt{2}\,F_\pi\,m_N\,\pi}~,& 
    d_p^{(\bar{p}p\eta_0)}\,&=\,-\,\frac{e\,(2\,D\,+\,3\,\lambda)\,\bar{g}_{ \bar{p}p\eta_0}\,m_{\eta_0}}{24\,\sqrt{2}\,F_\pi\,m_N\,\pi}~,
    \end{align}
    \begin{align}
    d_p^{(n\bar{p}\pi^+)}\,&=\,-\,\frac{e\,(D\,+\,F)\,\bar{g}_{n\bar{p}\pi^+}\,(3\pi\,m_\pi\,-\,2\,m_N\,\log(m_\pi^2/m_N^2))}{16\,\sqrt{2}\,F_\pi\,m_N\,\pi^2}~,\\
    d_p^{(\bar{p}\Lambda^0 K^+)}\,&=-\,\frac{e\,(D\,+\,3\,F)\,\bar{g}_{\bar{p}\Lambda^0 K^+}\,(3\pi\,m_K\,-\,2\,m_N\,\log(m_K^2/m_N^2))}{96\,\sqrt{2}\,F_\pi\,m_N\,\pi^2}~,\\
    d_p^{(\Sigma^0\bar{p}K^+)}\,&=-\,\frac{e\,(D\,-\,F)\,\bar{g}_{\Sigma^0\bar{p}K^+}\,(3\pi\,m_K\,-\,2\,m_N\,\log(m_K^2/m_N^2))}{32\,\sqrt{2}\,F_\pi\,m_N\,\pi^2}~.
\end{align}

For the couplings in terms of phases, we find: 
\begin{align}
    \bar{g}_{\bar{n}p\pi^-}&=-\frac{4\sqrt{2}\,B_0\,b_+(m_d\alpha_d+m_u\alpha_u)}{F_\pi}\,,\qquad\quad 
    \bar{g}_{\bar{n}\Sigma^- K^+}=-\frac{4\sqrt{2}\,B_0\,b_-(m_u\alpha_u+m_s\alpha_s)}{F_\pi}\,,
\end{align}
and
\begin{align}
        \bar{g}_{n\bar{p}\pi^+}&=-\frac{4\sqrt{2}B_0b_+(m_d\alpha_d+m_u\alpha_u)}{F_\pi}\,,& 
    \bar{g}_{\Sigma^+\bar{p}K^0}&=-\frac{4\sqrt{2}B_0b_-(m_d\alpha_d+m_s\alpha_s)}{F_\pi}\,,\\
    \bar{g}_{\bar{p}\Lambda^0 K^+}&=-\frac{4B_0(b_D+3b_F)(m_s\alpha_s+m_u\alpha_u)}{\sqrt{3}F_\pi}~,&
    \bar{g}_{\Sigma^0\bar{p}K^+}&=-\frac{4\sqrt{2}B_0b_-(m_s\alpha_s+m_u\alpha_u)}{F_\pi}\,,\\
    \bar{g}_{\bar{p}p\pi^0}&=\frac{8B_0[b_0(m_d\alpha_d-m_u\alpha_u) -b_+m_u\alpha_u]}{F_\pi}\,,&
    \bar{g}_{nn\pi^0}&=\frac{8B_0[b_0(m_d\alpha_d-m_u\alpha_u) +b_+m_d\alpha_d]}{F_\pi}\,,
\end{align}
\vspace*{-3ex}
\begin{align} 
    \bar{g}_{ \bar{p}p\eta_8}&=\frac{8B_0\big[b_0(2m_s\alpha_s-m_u\alpha_u-m_d\alpha_d)+2(b_D-b_F)m_s\alpha_s-(b_D+b_F)m_u\alpha_u\big]}{\sqrt{3}F_\pi}\,,\\
    \bar{g}_{ \bar{p}p\eta_0}&=-\frac{8\sqrt{\frac{2}{3}}B_0\big[b_0(m_u\alpha_u+m_d\alpha_d+m_s\alpha_s)+(b_D-b_F)m_s\alpha_s+(b_D+b_F)m_u\alpha_u\big]}{\sqrt{3}F_\pi} \, . 
\end{align}
In the above formulae, we have taken the 
large $N_C$-limit $F_\pi\simeq F_0$, with $F_\pi$ and $F_0$ being the pion and the eta decay constants, respectively. 

\smallskip

Taking these results into account, we obtain the relevant EDMs in terms of generic phases:
\begin{align} d_p&=\left\lbrace\alpha_u\left(-1.72^{+0.19}_{-0.24}\right)\times10^{-15}+\alpha_d\left(-2.63^{+0.14}_{-0.16}\right)\times10^{-15}+\alpha_s\left(-1.31^{+0.25}_{-0.30}\right)\times10^{-14}\right\rbrace\, e\,\text{cm},\label{eq:palpha}\\ 
d_n&=\left\lbrace
  \alpha_u\left(6.7^{+0.7}_{-0.9}\right)\times10^{-16}+
\alpha_d\left(10.6^{+0.6}_{-0.7}\right)\times10^{-16}+
\alpha_s\left(7.1^{+0.5}_{-0.5}\right)\times10^{-15}\right\rbrace\, e\,\text{cm}\,,\label{eq:nalpha}
\end{align}
where hadronic uncertainties from LECs and quark masses have been included. By considering the impact of the neutron and proton EDMs on the Mercury EDM~\cite{Dekens:2022gha}, we find
\begin{align}\label{eq:Hgalpha}
d_{\text{Hg}}=\left\lbrace
\alpha_u\left(-2.0^{+0.6}_{-0.7}\right)\times10^{-19}+
\alpha_d\left(-3.1^{+1.1}_{-1.2}\right)\times10^{-19}+
\alpha_s\left(-23\pm11\right)\times10^{-19}\right\rbrace\, e\,\text{cm}\,.
\end{align}
The neutron and Mercury EDM expressions should be compared with the experimental bounds, respectively $d_n^<=1.8 \times 10^{-26}\,e\,\text{cm}$
~\cite{Abel:2020pzs} and
 $ d_{\text{Hg}}^<=6.3 \times 10^{-30}\,e\,\text{cm}$~\cite{Graner:2016ses}.

The ThO EDM bound finally amounts to 
$\bar{g}_{\pi NN}^{(1)} \lesssim 4 \times 10^{-10}$ \cite{Flambaum:2019ejc}, 
with
\begin{equation}
\label{eq:ThO}    \bar{g}_{\pi NN}^{(1)}\equiv\frac12 (\bar g_{nn\pi}+\bar g_{pp\pi})=\frac{4\,B_0 (2\,b_0+b_D+b_F)(\alpha_d\,m_d-\alpha_u\,m_u)}{F_\pi} =
    \left(0.42^{+0.05}_{-0.07}\right)\alpha_u+\left(-0.89^{+0.04}_{-0.05}\right) \alpha_d\,.
\end{equation}
Eqs.~\eqref{eq:nalpha}--\eqref{eq:ThO} translate into  the 
constraints of \eqs{eq:dn}{eq:dtho} in the main text. 

\smallskip

We finally look for $\alpha_{u,d,s}$ that, while respecting these constraints,  try to  maximize the  CPV coupling in \eq{eq:gaN}, which numerically is 
\begin{align}
g_{aX}^S&= 10^{-11} \frac{10^9\,\GeV}{f_a}\big[(2.6^{+0.4}_{-0.5})\,\alpha_{u} + (2.6^{+0.4}_{-0.5})\,\alpha_{d}+(1.50^{+0.21}_{-0.27})\,\alpha_{s} \big]
    \,.
\end{align}
We find that all $\alpha_{u,d,s}$ are constrained to be of order $O(10^{-{9\div 10}})$, and lead to a maximal value  
of 
\begin{equation}
    \left|\frac{g_{aX}^{S}}{g_{aX}^{S,\theta}}\right| 
    \lesssim 10 \,,
\end{equation}
allowing also for the uncertainties in \eqs{eq:nalpha}{eq:ThO}.

\newpage
\twocolumngrid


\begin{thebibliography}{78}%
	\makeatletter
	\providecommand \@ifxundefined [1]{%
		\@ifx{#1\undefined}
	}%
	\providecommand \@ifnum [1]{%
		\ifnum #1\expandafter \@firstoftwo
		\else \expandafter \@secondoftwo
		\fi
	}%
	\providecommand \@ifx [1]{%
		\ifx #1\expandafter \@firstoftwo
		\else \expandafter \@secondoftwo
		\fi
	}%
	\providecommand \natexlab [1]{#1}%
	\providecommand \enquote  [1]{``#1''}%
	\providecommand \bibnamefont  [1]{#1}%
	\providecommand \bibfnamefont [1]{#1}%
	\providecommand \citenamefont [1]{#1}%
	\providecommand \href@noop [0]{\@secondoftwo}%
	\providecommand \href [0]{\begingroup \@sanitize@url \@href}%
	\providecommand \@href[1]{\@@startlink{#1}\@@href}%
	\providecommand \@@href[1]{\endgroup#1\@@endlink}%
	\providecommand \@sanitize@url [0]{\catcode `\\12\catcode `\$12\catcode
		`\&12\catcode `\#12\catcode `\^12\catcode `\_12\catcode `\%12\relax}%
	\providecommand \@@startlink[1]{}%
	\providecommand \@@endlink[0]{}%
	\providecommand \url  [0]{\begingroup\@sanitize@url \@url }%
	\providecommand \@url [1]{\endgroup\@href {#1}{\urlprefix }}%
	\providecommand \urlprefix  [0]{URL }%
	\providecommand \Eprint [0]{\href }%
	\providecommand \doibase [0]{http://dx.doi.org/}%
	\providecommand \selectlanguage [0]{\@gobble}%
	\providecommand \bibinfo  [0]{\@secondoftwo}%
	\providecommand \bibfield  [0]{\@secondoftwo}%
	\providecommand \translation [1]{[#1]}%
	\providecommand \BibitemOpen [0]{}%
	\providecommand \bibitemStop [0]{}%
	\providecommand \bibitemNoStop [0]{.\EOS\space}%
	\providecommand \EOS [0]{\spacefactor3000\relax}%
	\providecommand \BibitemShut  [1]{\csname bibitem#1\endcsname}%
	\let\auto@bib@innerbib\@empty
	%</preamble>
	\bibitem [{\citenamefont {Peccei}\ and\ \citenamefont
		{Quinn}(1977{\natexlab{a}})}]{Peccei:1977hh}%
	\BibitemOpen
	\bibfield  {author} {\bibinfo {author} {\bibfnamefont {R.~D.}\ \bibnamefont
			{Peccei}}\ and\ \bibinfo {author} {\bibfnamefont {H.~R.}\ \bibnamefont
			{Quinn}},\ }\href {\doibase 10.1103/PhysRevLett.38.1440} {\bibfield
		{journal} {\bibinfo  {journal} {Phys. Rev. Lett.}\ }\textbf {\bibinfo
			{volume} {38}},\ \bibinfo {pages} {1440} (\bibinfo {year}
		{1977}{\natexlab{a}})}\BibitemShut {NoStop}%
	\bibitem [{\citenamefont {Peccei}\ and\ \citenamefont
		{Quinn}(1977{\natexlab{b}})}]{Peccei:1977ur}%
	\BibitemOpen
	\bibfield  {author} {\bibinfo {author} {\bibfnamefont {R.~D.}\ \bibnamefont
			{Peccei}}\ and\ \bibinfo {author} {\bibfnamefont {H.~R.}\ \bibnamefont
			{Quinn}},\ }\href {\doibase 10.1103/PhysRevD.16.1791} {\bibfield  {journal}
		{\bibinfo  {journal} {Phys. Rev. D}\ }\textbf {\bibinfo {volume} {16}},\
		\bibinfo {pages} {1791} (\bibinfo {year} {1977}{\natexlab{b}})}\BibitemShut
	{NoStop}%
	\bibitem [{\citenamefont {Weinberg}(1978)}]{Weinberg:1977ma}%
	\BibitemOpen
	\bibfield  {author} {\bibinfo {author} {\bibfnamefont {S.}~\bibnamefont
			{Weinberg}},\ }\href {\doibase 10.1103/PhysRevLett.40.223} {\bibfield
		{journal} {\bibinfo  {journal} {Phys. Rev. Lett.}\ }\textbf {\bibinfo
			{volume} {40}},\ \bibinfo {pages} {223} (\bibinfo {year} {1978})}\BibitemShut
	{NoStop}%
	\bibitem [{\citenamefont {Wilczek}(1978)}]{Wilczek:1977pj}%
	\BibitemOpen
	\bibfield  {author} {\bibinfo {author} {\bibfnamefont {F.}~\bibnamefont
			{Wilczek}},\ }\href {\doibase 10.1103/PhysRevLett.40.279} {\bibfield
		{journal} {\bibinfo  {journal} {Phys. Rev. Lett.}\ }\textbf {\bibinfo
			{volume} {40}},\ \bibinfo {pages} {279} (\bibinfo {year} {1978})}\BibitemShut
	{NoStop}%
	\bibitem [{\citenamefont {Abel}\ \emph {et~al.}(2020)\citenamefont {Abel} \emph
		{et~al.}}]{Abel:2020pzs}%
	\BibitemOpen
	\bibfield  {author} {\bibinfo {author} {\bibfnamefont {C.}~\bibnamefont
			{Abel}} \emph {et~al.},\ }\href {\doibase 10.1103/PhysRevLett.124.081803}
	{\bibfield  {journal} {\bibinfo  {journal} {Phys. Rev. Lett.}\ }\textbf
		{\bibinfo {volume} {124}},\ \bibinfo {pages} {081803} (\bibinfo {year}
		{2020})},\ \Eprint {http://arxiv.org/abs/2001.11966} {arXiv:2001.11966
		[hep-ex]} \BibitemShut {NoStop}%
	\bibitem [{\citenamefont {Vafa}\ and\ \citenamefont
		{Witten}(1984)}]{Vafa:1984xg}%
	\BibitemOpen
	\bibfield  {author} {\bibinfo {author} {\bibfnamefont {C.}~\bibnamefont
			{Vafa}}\ and\ \bibinfo {author} {\bibfnamefont {E.}~\bibnamefont {Witten}},\
	}\href {\doibase 10.1103/PhysRevLett.53.535} {\bibfield  {journal} {\bibinfo
			{journal} {Phys. Rev. Lett.}\ }\textbf {\bibinfo {volume} {53}},\ \bibinfo
		{pages} {535} (\bibinfo {year} {1984})}\BibitemShut {NoStop}%
	\bibitem [{\citenamefont {Georgi}\ and\ \citenamefont
		{Randall}(1986)}]{Georgi:1986kr}%
	\BibitemOpen
	\bibfield  {author} {\bibinfo {author} {\bibfnamefont {H.}~\bibnamefont
			{Georgi}}\ and\ \bibinfo {author} {\bibfnamefont {L.}~\bibnamefont
			{Randall}},\ }\href {\doibase 10.1016/0550-3213(86)90022-2} {\bibfield
		{journal} {\bibinfo  {journal} {Nucl. Phys.}\ }\textbf {\bibinfo {volume}
			{B276}},\ \bibinfo {pages} {241} (\bibinfo {year} {1986})}\BibitemShut
	{NoStop}%
	%%CITATION = NUPHA,B276,241;%%
	\bibitem [{\citenamefont {Ellis}\ and\ \citenamefont
		{Gaillard}(1979)}]{Ellis:1978hq}%
	\BibitemOpen
	\bibfield  {author} {\bibinfo {author} {\bibfnamefont {J.~R.}\ \bibnamefont
			{Ellis}}\ and\ \bibinfo {author} {\bibfnamefont {M.~K.}\ \bibnamefont
			{Gaillard}},\ }\href {\doibase 10.1016/0550-3213(79)90297-9} {\bibfield
		{journal} {\bibinfo  {journal} {Nucl. Phys.}\ }\textbf {\bibinfo {volume}
			{B150}},\ \bibinfo {pages} {141} (\bibinfo {year} {1979})}\BibitemShut
	{NoStop}%
	%%CITATION = NUPHA,B150,141;%%
	\bibitem [{\citenamefont {Khriplovich}(1986)}]{Khriplovich:1985jr}%
	\BibitemOpen
	\bibfield  {author} {\bibinfo {author} {\bibfnamefont {I.~B.}\ \bibnamefont
			{Khriplovich}},\ }\href {\doibase 10.1016/0370-2693(86)90245-5} {\bibfield
		{journal} {\bibinfo  {journal} {Phys. Lett. B}\ }\textbf {\bibinfo {volume}
			{173}},\ \bibinfo {pages} {193} (\bibinfo {year} {1986})}\BibitemShut
	{NoStop}%
	\bibitem [{\citenamefont {G\'erard}\ and\ \citenamefont
		{Mertens}(2012)}]{Gerard:2012ud}%
	\BibitemOpen
	\bibfield  {author} {\bibinfo {author} {\bibfnamefont {J.-M.}\ \bibnamefont
			{G\'erard}}\ and\ \bibinfo {author} {\bibfnamefont {P.}~\bibnamefont
			{Mertens}},\ }\href {\doibase 10.1016/j.physletb.2012.08.036} {\bibfield
		{journal} {\bibinfo  {journal} {Phys. Lett. B}\ }\textbf {\bibinfo {volume}
			{716}},\ \bibinfo {pages} {316} (\bibinfo {year} {2012})},\ \Eprint
	{http://arxiv.org/abs/1206.0914} {arXiv:1206.0914 [hep-ph]} \BibitemShut
	{NoStop}%
	\bibitem [{\citenamefont {Kallosh}\ \emph {et~al.}(1995)\citenamefont
		{Kallosh}, \citenamefont {Linde}, \citenamefont {Linde},\ and\ \citenamefont
		{Susskind}}]{Kallosh:1995hi}%
	\BibitemOpen
	\bibfield  {author} {\bibinfo {author} {\bibfnamefont {R.}~\bibnamefont
			{Kallosh}}, \bibinfo {author} {\bibfnamefont {A.~D.}\ \bibnamefont {Linde}},
		\bibinfo {author} {\bibfnamefont {D.~A.}\ \bibnamefont {Linde}}, \ and\
		\bibinfo {author} {\bibfnamefont {L.}~\bibnamefont {Susskind}},\ }\href
	{\doibase 10.1103/PhysRevD.52.912} {\bibfield  {journal} {\bibinfo  {journal}
			{Phys. Rev. D}\ }\textbf {\bibinfo {volume} {52}},\ \bibinfo {pages} {912}
		(\bibinfo {year} {1995})},\ \Eprint {http://arxiv.org/abs/hep-th/9502069}
	{arXiv:hep-th/9502069} \BibitemShut {NoStop}%
	\bibitem [{\citenamefont {Moody}\ and\ \citenamefont
		{Wilczek}(1984)}]{Moody:1984ba}%
	\BibitemOpen
	\bibfield  {author} {\bibinfo {author} {\bibfnamefont {J.~E.}\ \bibnamefont
			{Moody}}\ and\ \bibinfo {author} {\bibfnamefont {F.}~\bibnamefont
			{Wilczek}},\ }\href {\doibase 10.1103/PhysRevD.30.130} {\bibfield  {journal}
		{\bibinfo  {journal} {Phys. Rev.}\ }\textbf {\bibinfo {volume} {D30}},\
		\bibinfo {pages} {130} (\bibinfo {year} {1984})}\BibitemShut {NoStop}%
	%%CITATION = PHRVA,D30,130;%%
	\bibitem [{\citenamefont {O'Hare}\ and\ \citenamefont
		{Vitagliano}(2020)}]{OHare:2020wah}%
	\BibitemOpen
	\bibfield  {author} {\bibinfo {author} {\bibfnamefont {C.~A.~J.}\
			\bibnamefont {O'Hare}}\ and\ \bibinfo {author} {\bibfnamefont
			{E.}~\bibnamefont {Vitagliano}},\ }\href {\doibase
		10.1103/PhysRevD.102.115026} {\bibfield  {journal} {\bibinfo  {journal}
			{Phys. Rev. D}\ }\textbf {\bibinfo {volume} {102}},\ \bibinfo {pages}
		{115026} (\bibinfo {year} {2020})},\ \Eprint
	{http://arxiv.org/abs/2010.03889} {arXiv:2010.03889 [hep-ph]} \BibitemShut
	{NoStop}%
	\bibitem [{\citenamefont {Raffelt}(2012)}]{Raffelt:2012sp}%
	\BibitemOpen
	\bibfield  {author} {\bibinfo {author} {\bibfnamefont {G.}~\bibnamefont
			{Raffelt}},\ }\href {\doibase 10.1103/PhysRevD.86.015001} {\bibfield
		{journal} {\bibinfo  {journal} {Phys. Rev. D}\ }\textbf {\bibinfo {volume}
			{86}},\ \bibinfo {pages} {015001} (\bibinfo {year} {2012})},\ \Eprint
	{http://arxiv.org/abs/1205.1776} {arXiv:1205.1776 [hep-ph]} \BibitemShut
	{NoStop}%
	\bibitem [{\citenamefont {Irastorza}\ and\ \citenamefont
		{Redondo}(2018)}]{Irastorza:2018dyq}%
	\BibitemOpen
	\bibfield  {author} {\bibinfo {author} {\bibfnamefont {I.~G.}\ \bibnamefont
			{Irastorza}}\ and\ \bibinfo {author} {\bibfnamefont {J.}~\bibnamefont
			{Redondo}},\ }\href {\doibase 10.1016/j.ppnp.2018.05.003} {\bibfield
		{journal} {\bibinfo  {journal} {Prog. Part. Nucl. Phys.}\ }\textbf {\bibinfo
			{volume} {102}},\ \bibinfo {pages} {89} (\bibinfo {year} {2018})},\ \Eprint
	{http://arxiv.org/abs/1801.08127} {arXiv:1801.08127 [hep-ph]} \BibitemShut
	{NoStop}%
	\bibitem [{\citenamefont {Sikivie}(2021)}]{Sikivie:2020zpn}%
	\BibitemOpen
	\bibfield  {author} {\bibinfo {author} {\bibfnamefont {P.}~\bibnamefont
			{Sikivie}},\ }\href {\doibase 10.1103/RevModPhys.93.015004} {\bibfield
		{journal} {\bibinfo  {journal} {Rev. Mod. Phys.}\ }\textbf {\bibinfo {volume}
			{93}},\ \bibinfo {pages} {015004} (\bibinfo {year} {2021})},\ \Eprint
	{http://arxiv.org/abs/2003.02206} {arXiv:2003.02206 [hep-ph]} \BibitemShut
	{NoStop}%
	\bibitem [{\citenamefont {Pospelov}(1998)}]{Pospelov:1997uv}%
	\BibitemOpen
	\bibfield  {author} {\bibinfo {author} {\bibfnamefont {M.}~\bibnamefont
			{Pospelov}},\ }\href {\doibase 10.1103/PhysRevD.58.097703} {\bibfield
		{journal} {\bibinfo  {journal} {Phys. Rev. D}\ }\textbf {\bibinfo {volume}
			{58}},\ \bibinfo {pages} {097703} (\bibinfo {year} {1998})},\ \Eprint
	{http://arxiv.org/abs/hep-ph/9707431} {arXiv:hep-ph/9707431} \BibitemShut
	{NoStop}%
	\bibitem [{\citenamefont {Bertolini}\ \emph {et~al.}(2021)\citenamefont
		{Bertolini}, \citenamefont {Di~Luzio},\ and\ \citenamefont
		{Nesti}}]{Bertolini:2020hjc}%
	\BibitemOpen
	\bibfield  {author} {\bibinfo {author} {\bibfnamefont {S.}~\bibnamefont
			{Bertolini}}, \bibinfo {author} {\bibfnamefont {L.}~\bibnamefont {Di~Luzio}},
		\ and\ \bibinfo {author} {\bibfnamefont {F.}~\bibnamefont {Nesti}},\ }\href
	{\doibase 10.1103/PhysRevLett.126.081801} {\bibfield  {journal} {\bibinfo
			{journal} {Phys. Rev. Lett.}\ }\textbf {\bibinfo {volume} {126}},\ \bibinfo
		{pages} {081801} (\bibinfo {year} {2021})},\ \Eprint
	{http://arxiv.org/abs/2006.12508} {arXiv:2006.12508 [hep-ph]} \BibitemShut
	{NoStop}%
	\bibitem [{\citenamefont {Okawa}\ \emph {et~al.}(2022)\citenamefont {Okawa},
		\citenamefont {Pospelov},\ and\ \citenamefont {Ritz}}]{Okawa:2021fto}%
	\BibitemOpen
	\bibfield  {author} {\bibinfo {author} {\bibfnamefont {S.}~\bibnamefont
			{Okawa}}, \bibinfo {author} {\bibfnamefont {M.}~\bibnamefont {Pospelov}}, \
		and\ \bibinfo {author} {\bibfnamefont {A.}~\bibnamefont {Ritz}},\ }\href
	{\doibase 10.1103/PhysRevD.105.075003} {\bibfield  {journal} {\bibinfo
			{journal} {Phys. Rev. D}\ }\textbf {\bibinfo {volume} {105}},\ \bibinfo
		{pages} {075003} (\bibinfo {year} {2022})},\ \Eprint
	{http://arxiv.org/abs/2111.08040} {arXiv:2111.08040 [hep-ph]} \BibitemShut
	{NoStop}%
	\bibitem [{\citenamefont {Dekens}\ \emph {et~al.}(2022)\citenamefont {Dekens},
		\citenamefont {de~Vries},\ and\ \citenamefont {Shain}}]{Dekens:2022gha}%
	\BibitemOpen
	\bibfield  {author} {\bibinfo {author} {\bibfnamefont {W.}~\bibnamefont
			{Dekens}}, \bibinfo {author} {\bibfnamefont {J.}~\bibnamefont {de~Vries}}, \
		and\ \bibinfo {author} {\bibfnamefont {S.}~\bibnamefont {Shain}},\ }\href
	{\doibase 10.1007/JHEP07(2022)014} {\bibfield  {journal} {\bibinfo  {journal}
			{JHEP}\ }\textbf {\bibinfo {volume} {07}},\ \bibinfo {pages} {014} (\bibinfo
		{year} {2022})},\ \Eprint {http://arxiv.org/abs/2203.11230} {arXiv:2203.11230
		[hep-ph]} \BibitemShut {NoStop}%
	\bibitem [{\citenamefont {Plakkot}\ \emph {et~al.}(2023)\citenamefont
		{Plakkot}, \citenamefont {Dekens}, \citenamefont {de~Vries},\ and\
		\citenamefont {Shain}}]{Plakkot:2023pui}%
	\BibitemOpen
	\bibfield  {author} {\bibinfo {author} {\bibfnamefont {V.}~\bibnamefont
			{Plakkot}}, \bibinfo {author} {\bibfnamefont {W.}~\bibnamefont {Dekens}},
		\bibinfo {author} {\bibfnamefont {J.}~\bibnamefont {de~Vries}}, \ and\
		\bibinfo {author} {\bibfnamefont {S.}~\bibnamefont {Shain}},\ }\href
	{\doibase 10.1007/JHEP11(2023)012} {\bibfield  {journal} {\bibinfo  {journal}
			{JHEP}\ }\textbf {\bibinfo {volume} {11}},\ \bibinfo {pages} {012} (\bibinfo
		{year} {2023})},\ \Eprint {http://arxiv.org/abs/2306.07065} {arXiv:2306.07065
		[hep-ph]} \BibitemShut {NoStop}%
	\bibitem [{\citenamefont {Di~Luzio}\ \emph {et~al.}(2024)\citenamefont
		{Di~Luzio}, \citenamefont {Levati},\ and\ \citenamefont
		{Paradisi}}]{DiLuzio:2023cuk}%
	\BibitemOpen
	\bibfield  {author} {\bibinfo {author} {\bibfnamefont {L.}~\bibnamefont
			{Di~Luzio}}, \bibinfo {author} {\bibfnamefont {G.}~\bibnamefont {Levati}}, \
		and\ \bibinfo {author} {\bibfnamefont {P.}~\bibnamefont {Paradisi}},\ }\href
	{\doibase 10.1007/JHEP02(2024)020} {\bibfield  {journal} {\bibinfo  {journal}
			{JHEP}\ }\textbf {\bibinfo {volume} {2024}},\ \bibinfo {pages} {020}
		(\bibinfo {year} {2024})},\ \Eprint {http://arxiv.org/abs/2311.12158}
	{arXiv:2311.12158 [hep-ph]} \BibitemShut {NoStop}%
	\bibitem [{\citenamefont {Di~Luzio}\ \emph {et~al.}(2023)\citenamefont
		{Di~Luzio}, \citenamefont {Gisbert}, \citenamefont {Levati}, \citenamefont
		{Paradisi},\ and\ \citenamefont {S\o{}rensen}}]{DiLuzio:2023lmd}%
	\BibitemOpen
	\bibfield  {author} {\bibinfo {author} {\bibfnamefont {L.}~\bibnamefont
			{Di~Luzio}}, \bibinfo {author} {\bibfnamefont {H.}~\bibnamefont {Gisbert}},
		\bibinfo {author} {\bibfnamefont {G.}~\bibnamefont {Levati}}, \bibinfo
		{author} {\bibfnamefont {P.}~\bibnamefont {Paradisi}}, \ and\ \bibinfo
		{author} {\bibfnamefont {P.}~\bibnamefont {S\o{}rensen}},\ }\href@noop {} {\
		(\bibinfo {year} {2023})},\ \Eprint {http://arxiv.org/abs/2312.17310}
	{arXiv:2312.17310 [hep-ph]} \BibitemShut {NoStop}%
	\bibitem [{\citenamefont {Zhitnitsky}(1980)}]{Zhitnitsky:1980tq}%
	\BibitemOpen
	\bibfield  {author} {\bibinfo {author} {\bibfnamefont {A.~R.}\ \bibnamefont
			{Zhitnitsky}},\ }\href@noop {} {\bibfield  {journal} {\bibinfo  {journal}
			{Sov. J. Nucl. Phys.}\ }\textbf {\bibinfo {volume} {31}},\ \bibinfo {pages}
		{260} (\bibinfo {year} {1980})}\BibitemShut {NoStop}%
	\bibitem [{\citenamefont {Dine}\ \emph {et~al.}(1981)\citenamefont {Dine},
		\citenamefont {Fischler},\ and\ \citenamefont {Srednicki}}]{Dine:1981rt}%
	\BibitemOpen
	\bibfield  {author} {\bibinfo {author} {\bibfnamefont {M.}~\bibnamefont
			{Dine}}, \bibinfo {author} {\bibfnamefont {W.}~\bibnamefont {Fischler}}, \
		and\ \bibinfo {author} {\bibfnamefont {M.}~\bibnamefont {Srednicki}},\ }\href
	{\doibase 10.1016/0370-2693(81)90590-6} {\bibfield  {journal} {\bibinfo
			{journal} {Phys. Lett. B}\ }\textbf {\bibinfo {volume} {104}},\ \bibinfo
		{pages} {199} (\bibinfo {year} {1981})}\BibitemShut {NoStop}%
	\bibitem [{\citenamefont {Di~Luzio}\ \emph {et~al.}(2020)\citenamefont
		{Di~Luzio}, \citenamefont {Giannotti}, \citenamefont {Nardi},\ and\
		\citenamefont {Visinelli}}]{DiLuzio:2020wdo}%
	\BibitemOpen
	\bibfield  {author} {\bibinfo {author} {\bibfnamefont {L.}~\bibnamefont
			{Di~Luzio}}, \bibinfo {author} {\bibfnamefont {M.}~\bibnamefont {Giannotti}},
		\bibinfo {author} {\bibfnamefont {E.}~\bibnamefont {Nardi}}, \ and\ \bibinfo
		{author} {\bibfnamefont {L.}~\bibnamefont {Visinelli}},\ }\href {\doibase
		10.1016/j.physrep.2020.06.002} {\bibfield  {journal} {\bibinfo  {journal}
			{Phys. Rept.}\ }\textbf {\bibinfo {volume} {870}},\ \bibinfo {pages} {1}
		(\bibinfo {year} {2020})},\ \Eprint {http://arxiv.org/abs/2003.01100}
	{arXiv:2003.01100 [hep-ph]} \BibitemShut {NoStop}%
	\bibitem [{\citenamefont {O'Hare}(2020)}]{AxionLimits}%
	\BibitemOpen
	\bibfield  {author} {\bibinfo {author} {\bibfnamefont {C.}~\bibnamefont
			{O'Hare}},\ }\href {\doibase 10.5281/zenodo.3932430} {\enquote {\bibinfo
			{title} {cajohare/axionlimits: Axionlimits},}\ }\bibinfo {howpublished}
	{\url{https://cajohare.github.io/AxionLimits/}} (\bibinfo {year}
	{2020})\BibitemShut {NoStop}%
	\bibitem [{\citenamefont {Pich}\ and\ \citenamefont
		{de~Rafael}(1991)}]{Pich:1991fq}%
	\BibitemOpen
	\bibfield  {author} {\bibinfo {author} {\bibfnamefont {A.}~\bibnamefont
			{Pich}}\ and\ \bibinfo {author} {\bibfnamefont {E.}~\bibnamefont
			{de~Rafael}},\ }\href {\doibase 10.1016/0550-3213(91)90019-T} {\bibfield
		{journal} {\bibinfo  {journal} {Nucl. Phys. B}\ }\textbf {\bibinfo {volume}
			{367}},\ \bibinfo {pages} {313} (\bibinfo {year} {1991})}\BibitemShut
	{NoStop}%
	\bibitem [{\citenamefont {Arvanitaki}\ and\ \citenamefont
		{Geraci}(2014)}]{Arvanitaki:2014dfa}%
	\BibitemOpen
	\bibfield  {author} {\bibinfo {author} {\bibfnamefont {A.}~\bibnamefont
			{Arvanitaki}}\ and\ \bibinfo {author} {\bibfnamefont {A.~A.}\ \bibnamefont
			{Geraci}},\ }\href {\doibase 10.1103/PhysRevLett.113.161801} {\bibfield
		{journal} {\bibinfo  {journal} {Phys. Rev. Lett.}\ }\textbf {\bibinfo
			{volume} {113}},\ \bibinfo {pages} {161801} (\bibinfo {year} {2014})},\
	\Eprint {http://arxiv.org/abs/1403.1290} {arXiv:1403.1290 [hep-ph]}
	\BibitemShut {NoStop}%
	%%CITATION = ARXIV:1403.1290;%%
	\bibitem [{\citenamefont {Geraci}\ \emph {et~al.}(2018)\citenamefont {Geraci}
		\emph {et~al.}}]{ARIADNE:2017tdd}%
	\BibitemOpen
	\bibfield  {author} {\bibinfo {author} {\bibfnamefont {A.~A.}\ \bibnamefont
			{Geraci}} \emph {et~al.} (\bibinfo {collaboration} {ARIADNE}),\ }\href
	{\doibase 10.1007/978-3-319-92726-8_18} {\bibfield  {journal} {\bibinfo
			{journal} {Springer Proc. Phys.}\ }\textbf {\bibinfo {volume} {211}},\
		\bibinfo {pages} {151} (\bibinfo {year} {2018})},\ \Eprint
	{http://arxiv.org/abs/1710.05413} {arXiv:1710.05413 [astro-ph.IM]}
	\BibitemShut {NoStop}%
	\bibitem [{\citenamefont {Cirigliano}\ \emph {et~al.}(2017)\citenamefont
		{Cirigliano}, \citenamefont {Dekens}, \citenamefont {de~Vries},\ and\
		\citenamefont {Mereghetti}}]{Cirigliano:2016yhc}%
	\BibitemOpen
	\bibfield  {author} {\bibinfo {author} {\bibfnamefont {V.}~\bibnamefont
			{Cirigliano}}, \bibinfo {author} {\bibfnamefont {W.}~\bibnamefont {Dekens}},
		\bibinfo {author} {\bibfnamefont {J.}~\bibnamefont {de~Vries}}, \ and\
		\bibinfo {author} {\bibfnamefont {E.}~\bibnamefont {Mereghetti}},\ }\href
	{\doibase 10.1016/j.physletb.2017.01.037} {\bibfield  {journal} {\bibinfo
			{journal} {Phys. Lett. B}\ }\textbf {\bibinfo {volume} {767}},\ \bibinfo
		{pages} {1} (\bibinfo {year} {2017})},\ \Eprint
	{http://arxiv.org/abs/1612.03914} {arXiv:1612.03914 [hep-ph]} \BibitemShut
	{NoStop}%
	\bibitem [{\citenamefont {Bertolini}\ \emph {et~al.}(2020)\citenamefont
		{Bertolini}, \citenamefont {Maiezza},\ and\ \citenamefont
		{Nesti}}]{Bertolini:2019out}%
	\BibitemOpen
	\bibfield  {author} {\bibinfo {author} {\bibfnamefont {S.}~\bibnamefont
			{Bertolini}}, \bibinfo {author} {\bibfnamefont {A.}~\bibnamefont {Maiezza}},
		\ and\ \bibinfo {author} {\bibfnamefont {F.}~\bibnamefont {Nesti}},\ }\href
	{\doibase 10.1103/PhysRevD.101.035036} {\bibfield  {journal} {\bibinfo
			{journal} {Phys. Rev. D}\ }\textbf {\bibinfo {volume} {101}},\ \bibinfo
		{pages} {035036} (\bibinfo {year} {2020})},\ \Eprint
	{http://arxiv.org/abs/1911.09472} {arXiv:1911.09472 [hep-ph]} \BibitemShut
	{NoStop}%
	\bibitem [{\citenamefont {Graner}\ \emph {et~al.}(2016)\citenamefont {Graner},
		\citenamefont {Chen}, \citenamefont {Lindahl},\ and\ \citenamefont
		{Heckel}}]{Graner:2016ses}%
	\BibitemOpen
	\bibfield  {author} {\bibinfo {author} {\bibfnamefont {B.}~\bibnamefont
			{Graner}}, \bibinfo {author} {\bibfnamefont {Y.}~\bibnamefont {Chen}},
		\bibinfo {author} {\bibfnamefont {E.~G.}\ \bibnamefont {Lindahl}}, \ and\
		\bibinfo {author} {\bibfnamefont {B.~R.}\ \bibnamefont {Heckel}},\ }\href
	{\doibase 10.1103/PhysRevLett.116.161601} {\bibfield  {journal} {\bibinfo
			{journal} {Phys. Rev. Lett.}\ }\textbf {\bibinfo {volume} {116}},\ \bibinfo
		{pages} {161601} (\bibinfo {year} {2016})},\ \bibinfo {note} {[Erratum:
		Phys.Rev.Lett. 119, 119901 (2017)]},\ \Eprint
	{http://arxiv.org/abs/1601.04339} {arXiv:1601.04339 [physics.atom-ph]}
	\BibitemShut {NoStop}%
	\bibitem [{\citenamefont {Flambaum}\ \emph {et~al.}(2020)\citenamefont
		{Flambaum}, \citenamefont {Pospelov}, \citenamefont {Ritz},\ and\
		\citenamefont {Stadnik}}]{Flambaum:2019ejc}%
	\BibitemOpen
	\bibfield  {author} {\bibinfo {author} {\bibfnamefont {V.~V.}\ \bibnamefont
			{Flambaum}}, \bibinfo {author} {\bibfnamefont {M.}~\bibnamefont {Pospelov}},
		\bibinfo {author} {\bibfnamefont {A.}~\bibnamefont {Ritz}}, \ and\ \bibinfo
		{author} {\bibfnamefont {Y.~V.}\ \bibnamefont {Stadnik}},\ }\href {\doibase
		10.1103/PhysRevD.102.035001} {\bibfield  {journal} {\bibinfo  {journal}
			{Phys. Rev. D}\ }\textbf {\bibinfo {volume} {102}},\ \bibinfo {pages}
		{035001} (\bibinfo {year} {2020})},\ \Eprint
	{http://arxiv.org/abs/1912.13129} {arXiv:1912.13129 [hep-ph]} \BibitemShut
	{NoStop}%
	\bibitem [{\citenamefont {Andreev}\ \emph {et~al.}(2018)\citenamefont {Andreev}
		\emph {et~al.}}]{ACME:2018yjb}%
	\BibitemOpen
	\bibfield  {author} {\bibinfo {author} {\bibfnamefont {V.}~\bibnamefont
			{Andreev}} \emph {et~al.} (\bibinfo {collaboration} {ACME}),\ }\href
	{\doibase 10.1038/s41586-018-0599-8} {\bibfield  {journal} {\bibinfo
			{journal} {Nature}\ }\textbf {\bibinfo {volume} {562}},\ \bibinfo {pages}
		{355} (\bibinfo {year} {2018})}\BibitemShut {NoStop}%
	\bibitem [{\citenamefont {Bishof}\ \emph {et~al.}(2016)\citenamefont {Bishof}
		\emph {et~al.}}]{Bishof:2016uqx}%
	\BibitemOpen
	\bibfield  {author} {\bibinfo {author} {\bibfnamefont {M.}~\bibnamefont
			{Bishof}} \emph {et~al.},\ }\href {\doibase 10.1103/PhysRevC.94.025501}
	{\bibfield  {journal} {\bibinfo  {journal} {Phys. Rev. C}\ }\textbf {\bibinfo
			{volume} {94}},\ \bibinfo {pages} {025501} (\bibinfo {year} {2016})},\
	\Eprint {http://arxiv.org/abs/1606.04931} {arXiv:1606.04931 [nucl-ex]}
	\BibitemShut {NoStop}%
	\bibitem [{\citenamefont {Hudson}\ \emph {et~al.}(2011)\citenamefont {Hudson},
		\citenamefont {Kara}, \citenamefont {Smallman}, \citenamefont {Sauer},
		\citenamefont {Tarbutt},\ and\ \citenamefont {Hinds}}]{Hudson:2011zz}%
	\BibitemOpen
	\bibfield  {author} {\bibinfo {author} {\bibfnamefont {J.~J.}\ \bibnamefont
			{Hudson}}, \bibinfo {author} {\bibfnamefont {D.~M.}\ \bibnamefont {Kara}},
		\bibinfo {author} {\bibfnamefont {I.~J.}\ \bibnamefont {Smallman}}, \bibinfo
		{author} {\bibfnamefont {B.~E.}\ \bibnamefont {Sauer}}, \bibinfo {author}
		{\bibfnamefont {M.~R.}\ \bibnamefont {Tarbutt}}, \ and\ \bibinfo {author}
		{\bibfnamefont {E.~A.}\ \bibnamefont {Hinds}},\ }\href {\doibase
		10.1038/nature10104} {\bibfield  {journal} {\bibinfo  {journal} {Nature}\
		}\textbf {\bibinfo {volume} {473}},\ \bibinfo {pages} {493} (\bibinfo {year}
		{2011})}\BibitemShut {NoStop}%
	\bibitem [{\citenamefont {Cairncross}\ \emph {et~al.}(2017)\citenamefont
		{Cairncross}, \citenamefont {Gresh}, \citenamefont {Grau}, \citenamefont
		{Cossel}, \citenamefont {Roussy}, \citenamefont {Ni}, \citenamefont {Zhou},
		\citenamefont {Ye},\ and\ \citenamefont {Cornell}}]{Cairncross:2017fip}%
	\BibitemOpen
	\bibfield  {author} {\bibinfo {author} {\bibfnamefont {W.~B.}\ \bibnamefont
			{Cairncross}}, \bibinfo {author} {\bibfnamefont {D.~N.}\ \bibnamefont
			{Gresh}}, \bibinfo {author} {\bibfnamefont {M.}~\bibnamefont {Grau}},
		\bibinfo {author} {\bibfnamefont {K.~C.}\ \bibnamefont {Cossel}}, \bibinfo
		{author} {\bibfnamefont {T.~S.}\ \bibnamefont {Roussy}}, \bibinfo {author}
		{\bibfnamefont {Y.}~\bibnamefont {Ni}}, \bibinfo {author} {\bibfnamefont
			{Y.}~\bibnamefont {Zhou}}, \bibinfo {author} {\bibfnamefont {J.}~\bibnamefont
			{Ye}}, \ and\ \bibinfo {author} {\bibfnamefont {E.~A.}\ \bibnamefont
			{Cornell}},\ }\href {\doibase 10.1103/PhysRevLett.119.153001} {\bibfield
		{journal} {\bibinfo  {journal} {Phys. Rev. Lett.}\ }\textbf {\bibinfo
			{volume} {119}},\ \bibinfo {pages} {153001} (\bibinfo {year} {2017})},\
	\Eprint {http://arxiv.org/abs/1704.07928} {arXiv:1704.07928
		[physics.atom-ph]} \BibitemShut {NoStop}%
	\bibitem [{\citenamefont {Allwicher}\ \emph {et~al.}(2023)\citenamefont
		{Allwicher}, \citenamefont {Faroughy}, \citenamefont {Jaffredo},
		\citenamefont {Sumensari},\ and\ \citenamefont {Wilsch}}]{Allwicher:2022gkm}%
	\BibitemOpen
	\bibfield  {author} {\bibinfo {author} {\bibfnamefont {L.}~\bibnamefont
			{Allwicher}}, \bibinfo {author} {\bibfnamefont {D.~A.}\ \bibnamefont
			{Faroughy}}, \bibinfo {author} {\bibfnamefont {F.}~\bibnamefont {Jaffredo}},
		\bibinfo {author} {\bibfnamefont {O.}~\bibnamefont {Sumensari}}, \ and\
		\bibinfo {author} {\bibfnamefont {F.}~\bibnamefont {Wilsch}},\ }\href
	{\doibase 10.1007/JHEP03(2023)064} {\bibfield  {journal} {\bibinfo  {journal}
			{JHEP}\ }\textbf {\bibinfo {volume} {03}},\ \bibinfo {pages} {064} (\bibinfo
		{year} {2023})},\ \Eprint {http://arxiv.org/abs/2207.10714} {arXiv:2207.10714
		[hep-ph]} \BibitemShut {NoStop}%
	\bibitem [{\citenamefont {Zhang}(2023)}]{Zhang:2022ykd}%
	\BibitemOpen
	\bibfield  {author} {\bibinfo {author} {\bibfnamefont {Y.}~\bibnamefont
			{Zhang}},\ }\href {\doibase 10.1103/PhysRevD.107.055025} {\bibfield
		{journal} {\bibinfo  {journal} {Phys. Rev. D}\ }\textbf {\bibinfo {volume}
			{107}},\ \bibinfo {pages} {055025} (\bibinfo {year} {2023})},\ \Eprint
	{http://arxiv.org/abs/2209.09429} {arXiv:2209.09429 [hep-ph]} \BibitemShut
	{NoStop}%
	\bibitem [{\citenamefont {Zhang}(2024)}]{Zhang:2023gfu}%
	\BibitemOpen
	\bibfield  {author} {\bibinfo {author} {\bibfnamefont {Y.}~\bibnamefont
			{Zhang}},\ }\href {\doibase 10.1103/PhysRevLett.132.081003} {\bibfield
		{journal} {\bibinfo  {journal} {Phys. Rev. Lett.}\ }\textbf {\bibinfo
			{volume} {132}},\ \bibinfo {pages} {081003} (\bibinfo {year} {2024})},\
	\Eprint {http://arxiv.org/abs/2305.15495} {arXiv:2305.15495 [hep-ph]}
	\BibitemShut {NoStop}%
	\bibitem [{\citenamefont {Di~Luzio}\ and\ \citenamefont
		{S\o{}rensen}(2024)}]{DiLuzio:2024fyt}%
	\BibitemOpen
	\bibfield  {author} {\bibinfo {author} {\bibfnamefont {L.}~\bibnamefont
			{Di~Luzio}}\ and\ \bibinfo {author} {\bibfnamefont {P.}~\bibnamefont
			{S\o{}rensen}},\ }\href@noop {} {\  (\bibinfo {year} {2024})},\ \Eprint
	{http://arxiv.org/abs/2408.04623} {arXiv:2408.04623 [hep-ph]} \BibitemShut
	{NoStop}%
	\bibitem [{\citenamefont {Di~Luzio}\ \emph
		{et~al.}(2021{\natexlab{a}})\citenamefont {Di~Luzio}, \citenamefont
		{Gr\"ober},\ and\ \citenamefont {Paradisi}}]{DiLuzio:2020oah}%
	\BibitemOpen
	\bibfield  {author} {\bibinfo {author} {\bibfnamefont {L.}~\bibnamefont
			{Di~Luzio}}, \bibinfo {author} {\bibfnamefont {R.}~\bibnamefont {Gr\"ober}},
		\ and\ \bibinfo {author} {\bibfnamefont {P.}~\bibnamefont {Paradisi}},\
	}\href {\doibase 10.1103/PhysRevD.104.095027} {\bibfield  {journal} {\bibinfo
			{journal} {Phys. Rev. D}\ }\textbf {\bibinfo {volume} {104}},\ \bibinfo
		{pages} {095027} (\bibinfo {year} {2021}{\natexlab{a}})},\ \Eprint
	{http://arxiv.org/abs/2010.13760} {arXiv:2010.13760 [hep-ph]} \BibitemShut
	{NoStop}%
	\bibitem [{\citenamefont {Hook}(2018)}]{Hook:2018jle}%
	\BibitemOpen
	\bibfield  {author} {\bibinfo {author} {\bibfnamefont {A.}~\bibnamefont
			{Hook}},\ }\href {\doibase 10.1103/PhysRevLett.120.261802} {\bibfield
		{journal} {\bibinfo  {journal} {Phys. Rev. Lett.}\ }\textbf {\bibinfo
			{volume} {120}},\ \bibinfo {pages} {261802} (\bibinfo {year} {2018})},\
	\Eprint {http://arxiv.org/abs/1802.10093} {arXiv:1802.10093 [hep-ph]}
	\BibitemShut {NoStop}%
	\bibitem [{\citenamefont {Di~Luzio}\ \emph
		{et~al.}(2021{\natexlab{b}})\citenamefont {Di~Luzio}, \citenamefont {Gavela},
		\citenamefont {Quilez},\ and\ \citenamefont {Ringwald}}]{DiLuzio:2021pxd}%
	\BibitemOpen
	\bibfield  {author} {\bibinfo {author} {\bibfnamefont {L.}~\bibnamefont
			{Di~Luzio}}, \bibinfo {author} {\bibfnamefont {B.}~\bibnamefont {Gavela}},
		\bibinfo {author} {\bibfnamefont {P.}~\bibnamefont {Quilez}}, \ and\ \bibinfo
		{author} {\bibfnamefont {A.}~\bibnamefont {Ringwald}},\ }\href {\doibase
		10.1007/JHEP05(2021)184} {\bibfield  {journal} {\bibinfo  {journal} {JHEP}\
		}\textbf {\bibinfo {volume} {05}},\ \bibinfo {pages} {184} (\bibinfo {year}
		{2021}{\natexlab{b}})},\ \Eprint {http://arxiv.org/abs/2102.00012}
	{arXiv:2102.00012 [hep-ph]} \BibitemShut {NoStop}%
	\bibitem [{\citenamefont {Di~Luzio}\ \emph
		{et~al.}(2021{\natexlab{c}})\citenamefont {Di~Luzio}, \citenamefont {Gavela},
		\citenamefont {Quilez},\ and\ \citenamefont {Ringwald}}]{DiLuzio:2021gos}%
	\BibitemOpen
	\bibfield  {author} {\bibinfo {author} {\bibfnamefont {L.}~\bibnamefont
			{Di~Luzio}}, \bibinfo {author} {\bibfnamefont {B.}~\bibnamefont {Gavela}},
		\bibinfo {author} {\bibfnamefont {P.}~\bibnamefont {Quilez}}, \ and\ \bibinfo
		{author} {\bibfnamefont {A.}~\bibnamefont {Ringwald}},\ }\href {\doibase
		10.1088/1475-7516/2021/10/001} {\bibfield  {journal} {\bibinfo  {journal}
			{JCAP}\ }\textbf {\bibinfo {volume} {10}},\ \bibinfo {pages} {001} (\bibinfo
		{year} {2021}{\natexlab{c}})},\ \Eprint {http://arxiv.org/abs/2102.01082}
	{arXiv:2102.01082 [hep-ph]} \BibitemShut {NoStop}%
	\bibitem [{\citenamefont {Hook}\ and\ \citenamefont
		{Huang}(2018)}]{Hook:2017psm}%
	\BibitemOpen
	\bibfield  {author} {\bibinfo {author} {\bibfnamefont {A.}~\bibnamefont
			{Hook}}\ and\ \bibinfo {author} {\bibfnamefont {J.}~\bibnamefont {Huang}},\
	}\href {\doibase 10.1007/JHEP06(2018)036} {\bibfield  {journal} {\bibinfo
			{journal} {JHEP}\ }\textbf {\bibinfo {volume} {06}},\ \bibinfo {pages} {036}
		(\bibinfo {year} {2018})},\ \Eprint {http://arxiv.org/abs/1708.08464}
	{arXiv:1708.08464 [hep-ph]} \BibitemShut {NoStop}%
	\bibitem [{\citenamefont {Balkin}\ \emph {et~al.}(2024)\citenamefont {Balkin},
		\citenamefont {Serra}, \citenamefont {Springmann}, \citenamefont {Stelzl},\
		and\ \citenamefont {Weiler}}]{Balkin:2022qer}%
	\BibitemOpen
	\bibfield  {author} {\bibinfo {author} {\bibfnamefont {R.}~\bibnamefont
			{Balkin}}, \bibinfo {author} {\bibfnamefont {J.}~\bibnamefont {Serra}},
		\bibinfo {author} {\bibfnamefont {K.}~\bibnamefont {Springmann}}, \bibinfo
		{author} {\bibfnamefont {S.}~\bibnamefont {Stelzl}}, \ and\ \bibinfo {author}
		{\bibfnamefont {A.}~\bibnamefont {Weiler}},\ }\href {\doibase
		10.1103/PhysRevD.109.095032} {\bibfield  {journal} {\bibinfo  {journal}
			{Phys. Rev. D}\ }\textbf {\bibinfo {volume} {109}},\ \bibinfo {pages}
		{095032} (\bibinfo {year} {2024})},\ \Eprint
	{http://arxiv.org/abs/2211.02661} {arXiv:2211.02661 [hep-ph]} \BibitemShut
	{NoStop}%
	\bibitem [{\citenamefont {Chen}\ \emph {et~al.}(2016)\citenamefont {Chen},
		\citenamefont {Tham}, \citenamefont {Krause}, \citenamefont {Lopez},
		\citenamefont {Fischbach},\ and\ \citenamefont {Decca}}]{Chen:2014oda}%
	\BibitemOpen
	\bibfield  {author} {\bibinfo {author} {\bibfnamefont {Y.~J.}\ \bibnamefont
			{Chen}}, \bibinfo {author} {\bibfnamefont {W.~K.}\ \bibnamefont {Tham}},
		\bibinfo {author} {\bibfnamefont {D.~E.}\ \bibnamefont {Krause}}, \bibinfo
		{author} {\bibfnamefont {D.}~\bibnamefont {Lopez}}, \bibinfo {author}
		{\bibfnamefont {E.}~\bibnamefont {Fischbach}}, \ and\ \bibinfo {author}
		{\bibfnamefont {R.~S.}\ \bibnamefont {Decca}},\ }\href {\doibase
		10.1103/PhysRevLett.116.221102} {\bibfield  {journal} {\bibinfo  {journal}
			{Phys. Rev. Lett.}\ }\textbf {\bibinfo {volume} {116}},\ \bibinfo {pages}
		{221102} (\bibinfo {year} {2016})},\ \Eprint {http://arxiv.org/abs/1410.7267}
	{arXiv:1410.7267 [hep-ex]} \BibitemShut {NoStop}%
	\bibitem [{\citenamefont {Kapner}\ \emph {et~al.}(2007)\citenamefont {Kapner},
		\citenamefont {Cook}, \citenamefont {Adelberger}, \citenamefont {Gundlach},
		\citenamefont {Heckel}, \citenamefont {Hoyle},\ and\ \citenamefont
		{Swanson}}]{Kapner:2006si}%
	\BibitemOpen
	\bibfield  {author} {\bibinfo {author} {\bibfnamefont {D.~J.}\ \bibnamefont
			{Kapner}}, \bibinfo {author} {\bibfnamefont {T.~S.}\ \bibnamefont {Cook}},
		\bibinfo {author} {\bibfnamefont {E.~G.}\ \bibnamefont {Adelberger}},
		\bibinfo {author} {\bibfnamefont {J.~H.}\ \bibnamefont {Gundlach}}, \bibinfo
		{author} {\bibfnamefont {B.~R.}\ \bibnamefont {Heckel}}, \bibinfo {author}
		{\bibfnamefont {C.~D.}\ \bibnamefont {Hoyle}}, \ and\ \bibinfo {author}
		{\bibfnamefont {H.~E.}\ \bibnamefont {Swanson}},\ }\href {\doibase
		10.1103/PhysRevLett.98.021101} {\bibfield  {journal} {\bibinfo  {journal}
			{Phys. Rev. Lett.}\ }\textbf {\bibinfo {volume} {98}},\ \bibinfo {pages}
		{021101} (\bibinfo {year} {2007})},\ \Eprint
	{http://arxiv.org/abs/hep-ph/0611184} {arXiv:hep-ph/0611184} \BibitemShut
	{NoStop}%
	\bibitem [{\citenamefont {Lee}\ \emph {et~al.}(2020)\citenamefont {Lee},
		\citenamefont {Adelberger}, \citenamefont {Cook}, \citenamefont {Fleischer},\
		and\ \citenamefont {Heckel}}]{Lee:2020zjt}%
	\BibitemOpen
	\bibfield  {author} {\bibinfo {author} {\bibfnamefont {J.~G.}\ \bibnamefont
			{Lee}}, \bibinfo {author} {\bibfnamefont {E.~G.}\ \bibnamefont {Adelberger}},
		\bibinfo {author} {\bibfnamefont {T.~S.}\ \bibnamefont {Cook}}, \bibinfo
		{author} {\bibfnamefont {S.~M.}\ \bibnamefont {Fleischer}}, \ and\ \bibinfo
		{author} {\bibfnamefont {B.~R.}\ \bibnamefont {Heckel}},\ }\href {\doibase
		10.1103/PhysRevLett.124.101101} {\bibfield  {journal} {\bibinfo  {journal}
			{Phys. Rev. Lett.}\ }\textbf {\bibinfo {volume} {124}},\ \bibinfo {pages}
		{101101} (\bibinfo {year} {2020})},\ \Eprint
	{http://arxiv.org/abs/2002.11761} {arXiv:2002.11761 [hep-ex]} \BibitemShut
	{NoStop}%
	\bibitem [{\citenamefont {Geraci}\ \emph {et~al.}(2008)\citenamefont {Geraci},
		\citenamefont {Smullin}, \citenamefont {Weld}, \citenamefont {Chiaverini},\
		and\ \citenamefont {Kapitulnik}}]{Geraci:2008hb}%
	\BibitemOpen
	\bibfield  {author} {\bibinfo {author} {\bibfnamefont {A.~A.}\ \bibnamefont
			{Geraci}}, \bibinfo {author} {\bibfnamefont {S.~J.}\ \bibnamefont {Smullin}},
		\bibinfo {author} {\bibfnamefont {D.~M.}\ \bibnamefont {Weld}}, \bibinfo
		{author} {\bibfnamefont {J.}~\bibnamefont {Chiaverini}}, \ and\ \bibinfo
		{author} {\bibfnamefont {A.}~\bibnamefont {Kapitulnik}},\ }\href {\doibase
		10.1103/PhysRevD.78.022002} {\bibfield  {journal} {\bibinfo  {journal} {Phys.
				Rev. D}\ }\textbf {\bibinfo {volume} {78}},\ \bibinfo {pages} {022002}
		(\bibinfo {year} {2008})},\ \Eprint {http://arxiv.org/abs/0802.2350}
	{arXiv:0802.2350 [hep-ex]} \BibitemShut {NoStop}%
	\bibitem [{\citenamefont {Yang}\ \emph {et~al.}(2012)\citenamefont {Yang},
		\citenamefont {Zhan}, \citenamefont {Wang}, \citenamefont {Shao},
		\citenamefont {Tu}, \citenamefont {Tan},\ and\ \citenamefont
		{Luo}}]{Yang:2012zzb}%
	\BibitemOpen
	\bibfield  {author} {\bibinfo {author} {\bibfnamefont {S.-Q.}\ \bibnamefont
			{Yang}}, \bibinfo {author} {\bibfnamefont {B.-F.}\ \bibnamefont {Zhan}},
		\bibinfo {author} {\bibfnamefont {Q.-L.}\ \bibnamefont {Wang}}, \bibinfo
		{author} {\bibfnamefont {C.-G.}\ \bibnamefont {Shao}}, \bibinfo {author}
		{\bibfnamefont {L.-C.}\ \bibnamefont {Tu}}, \bibinfo {author} {\bibfnamefont
			{W.-H.}\ \bibnamefont {Tan}}, \ and\ \bibinfo {author} {\bibfnamefont
			{J.}~\bibnamefont {Luo}},\ }\href {\doibase 10.1103/PhysRevLett.108.081101}
	{\bibfield  {journal} {\bibinfo  {journal} {Phys. Rev. Lett.}\ }\textbf
		{\bibinfo {volume} {108}},\ \bibinfo {pages} {081101} (\bibinfo {year}
		{2012})}\BibitemShut {NoStop}%
	\bibitem [{\citenamefont {Tan}\ \emph {et~al.}(2020)\citenamefont {Tan} \emph
		{et~al.}}]{Tan:2020vpf}%
	\BibitemOpen
	\bibfield  {author} {\bibinfo {author} {\bibfnamefont {W.-H.}\ \bibnamefont
			{Tan}} \emph {et~al.},\ }\href {\doibase 10.1103/PhysRevLett.124.051301}
	{\bibfield  {journal} {\bibinfo  {journal} {Phys. Rev. Lett.}\ }\textbf
		{\bibinfo {volume} {124}},\ \bibinfo {pages} {051301} (\bibinfo {year}
		{2020})}\BibitemShut {NoStop}%
	\bibitem [{\citenamefont {Ke}\ \emph {et~al.}(2021)\citenamefont {Ke},
		\citenamefont {Luo}, \citenamefont {Shao}, \citenamefont {Tan}, \citenamefont
		{Tan},\ and\ \citenamefont {Yang}}]{Ke:2021jtj}%
	\BibitemOpen
	\bibfield  {author} {\bibinfo {author} {\bibfnamefont {J.}~\bibnamefont
			{Ke}}, \bibinfo {author} {\bibfnamefont {J.}~\bibnamefont {Luo}}, \bibinfo
		{author} {\bibfnamefont {C.-G.}\ \bibnamefont {Shao}}, \bibinfo {author}
		{\bibfnamefont {Y.-J.}\ \bibnamefont {Tan}}, \bibinfo {author} {\bibfnamefont
			{W.-H.}\ \bibnamefont {Tan}}, \ and\ \bibinfo {author} {\bibfnamefont
			{S.-Q.}\ \bibnamefont {Yang}},\ }\href {\doibase
		10.1103/PhysRevLett.126.211101} {\bibfield  {journal} {\bibinfo  {journal}
			{Phys. Rev. Lett.}\ }\textbf {\bibinfo {volume} {126}},\ \bibinfo {pages}
		{211101} (\bibinfo {year} {2021})}\BibitemShut {NoStop}%
	\bibitem [{\citenamefont {Smith}\ \emph {et~al.}(2000)\citenamefont {Smith},
		\citenamefont {Hoyle}, \citenamefont {Gundlach}, \citenamefont {Adelberger},
		\citenamefont {Heckel},\ and\ \citenamefont {Swanson}}]{Smith:1999cr}%
	\BibitemOpen
	\bibfield  {author} {\bibinfo {author} {\bibfnamefont {G.~L.}\ \bibnamefont
			{Smith}}, \bibinfo {author} {\bibfnamefont {C.~D.}\ \bibnamefont {Hoyle}},
		\bibinfo {author} {\bibfnamefont {J.~H.}\ \bibnamefont {Gundlach}}, \bibinfo
		{author} {\bibfnamefont {E.~G.}\ \bibnamefont {Adelberger}}, \bibinfo
		{author} {\bibfnamefont {B.~R.}\ \bibnamefont {Heckel}}, \ and\ \bibinfo
		{author} {\bibfnamefont {H.~E.}\ \bibnamefont {Swanson}},\ }\href {\doibase
		10.1103/PhysRevD.61.022001} {\bibfield  {journal} {\bibinfo  {journal} {Phys.
				Rev. D}\ }\textbf {\bibinfo {volume} {61}},\ \bibinfo {pages} {022001}
		(\bibinfo {year} {2000})},\ \Eprint {http://arxiv.org/abs/2405.10982}
	{arXiv:2405.10982 [gr-qc]} \BibitemShut {NoStop}%
	\bibitem [{\citenamefont {Berg\'e}\ \emph {et~al.}(2018)\citenamefont
		{Berg\'e}, \citenamefont {Brax}, \citenamefont {M\'etris}, \citenamefont
		{Pernot-Borr\`as}, \citenamefont {Touboul},\ and\ \citenamefont
		{Uzan}}]{Berge:2017ovy}%
	\BibitemOpen
	\bibfield  {author} {\bibinfo {author} {\bibfnamefont {J.}~\bibnamefont
			{Berg\'e}}, \bibinfo {author} {\bibfnamefont {P.}~\bibnamefont {Brax}},
		\bibinfo {author} {\bibfnamefont {G.}~\bibnamefont {M\'etris}}, \bibinfo
		{author} {\bibfnamefont {M.}~\bibnamefont {Pernot-Borr\`as}}, \bibinfo
		{author} {\bibfnamefont {P.}~\bibnamefont {Touboul}}, \ and\ \bibinfo
		{author} {\bibfnamefont {J.-P.}\ \bibnamefont {Uzan}},\ }\href {\doibase
		10.1103/PhysRevLett.120.141101} {\bibfield  {journal} {\bibinfo  {journal}
			{Phys. Rev. Lett.}\ }\textbf {\bibinfo {volume} {120}},\ \bibinfo {pages}
		{141101} (\bibinfo {year} {2018})},\ \Eprint
	{http://arxiv.org/abs/1712.00483} {arXiv:1712.00483 [gr-qc]} \BibitemShut
	{NoStop}%
	\bibitem [{\citenamefont {Crescini}\ \emph
		{et~al.}(2017{\natexlab{a}})\citenamefont {Crescini}, \citenamefont
		{Braggio}, \citenamefont {Carugno}, \citenamefont {Falferi}, \citenamefont
		{Ortolan},\ and\ \citenamefont {Ruoso}}]{Crescini:2016lwj}%
	\BibitemOpen
	\bibfield  {author} {\bibinfo {author} {\bibfnamefont {N.}~\bibnamefont
			{Crescini}}, \bibinfo {author} {\bibfnamefont {C.}~\bibnamefont {Braggio}},
		\bibinfo {author} {\bibfnamefont {G.}~\bibnamefont {Carugno}}, \bibinfo
		{author} {\bibfnamefont {P.}~\bibnamefont {Falferi}}, \bibinfo {author}
		{\bibfnamefont {A.}~\bibnamefont {Ortolan}}, \ and\ \bibinfo {author}
		{\bibfnamefont {G.}~\bibnamefont {Ruoso}},\ }\href {\doibase
		10.1016/j.nima.2016.10.050} {\bibfield  {journal} {\bibinfo  {journal} {Nucl.
				Instrum. Meth. A}\ }\textbf {\bibinfo {volume} {842}},\ \bibinfo {pages}
		{109} (\bibinfo {year} {2017}{\natexlab{a}})},\ \Eprint
	{http://arxiv.org/abs/1606.04751} {arXiv:1606.04751 [physics.ins-det]}
	\BibitemShut {NoStop}%
	\bibitem [{\citenamefont {Crescini}\ \emph
		{et~al.}(2017{\natexlab{b}})\citenamefont {Crescini}, \citenamefont
		{Braggio}, \citenamefont {Carugno}, \citenamefont {Falferi}, \citenamefont
		{Ortolan},\ and\ \citenamefont {Ruoso}}]{Crescini:2017uxs}%
	\BibitemOpen
	\bibfield  {author} {\bibinfo {author} {\bibfnamefont {N.}~\bibnamefont
			{Crescini}}, \bibinfo {author} {\bibfnamefont {C.}~\bibnamefont {Braggio}},
		\bibinfo {author} {\bibfnamefont {G.}~\bibnamefont {Carugno}}, \bibinfo
		{author} {\bibfnamefont {P.}~\bibnamefont {Falferi}}, \bibinfo {author}
		{\bibfnamefont {A.}~\bibnamefont {Ortolan}}, \ and\ \bibinfo {author}
		{\bibfnamefont {G.}~\bibnamefont {Ruoso}},\ }\href {\doibase
		10.1016/j.physletb.2017.09.019} {\bibfield  {journal} {\bibinfo  {journal}
			{Phys. Lett. B}\ }\textbf {\bibinfo {volume} {773}},\ \bibinfo {pages} {677}
		(\bibinfo {year} {2017}{\natexlab{b}})},\ \Eprint
	{http://arxiv.org/abs/1705.06044} {arXiv:1705.06044 [hep-ex]} \BibitemShut
	{NoStop}%
	\bibitem [{\citenamefont {Crescini}\ \emph {et~al.}(2022)\citenamefont
		{Crescini}, \citenamefont {Carugno}, \citenamefont {Falferi}, \citenamefont
		{Ortolan}, \citenamefont {Ruoso},\ and\ \citenamefont
		{Speake}}]{Crescini:2020ykp}%
	\BibitemOpen
	\bibfield  {author} {\bibinfo {author} {\bibfnamefont {N.}~\bibnamefont
			{Crescini}}, \bibinfo {author} {\bibfnamefont {G.}~\bibnamefont {Carugno}},
		\bibinfo {author} {\bibfnamefont {P.}~\bibnamefont {Falferi}}, \bibinfo
		{author} {\bibfnamefont {A.}~\bibnamefont {Ortolan}}, \bibinfo {author}
		{\bibfnamefont {G.}~\bibnamefont {Ruoso}}, \ and\ \bibinfo {author}
		{\bibfnamefont {C.~C.}\ \bibnamefont {Speake}},\ }\href {\doibase
		10.1103/PhysRevD.105.022007} {\bibfield  {journal} {\bibinfo  {journal}
			{Phys. Rev. D}\ }\textbf {\bibinfo {volume} {105}},\ \bibinfo {pages}
		{022007} (\bibinfo {year} {2022})},\ \Eprint
	{http://arxiv.org/abs/2011.07100} {arXiv:2011.07100 [hep-ex]} \BibitemShut
	{NoStop}%
	\bibitem [{\citenamefont {Wineland}\ \emph {et~al.}(1991)\citenamefont
		{Wineland}, \citenamefont {Bollinger}, \citenamefont {Heinzen}, \citenamefont
		{Itano},\ and\ \citenamefont {Raizen}}]{Wineland:1991zz}%
	\BibitemOpen
	\bibfield  {author} {\bibinfo {author} {\bibfnamefont {D.~J.}\ \bibnamefont
			{Wineland}}, \bibinfo {author} {\bibfnamefont {J.~J.}\ \bibnamefont
			{Bollinger}}, \bibinfo {author} {\bibfnamefont {D.~J.}\ \bibnamefont
			{Heinzen}}, \bibinfo {author} {\bibfnamefont {W.~M.}\ \bibnamefont {Itano}},
		\ and\ \bibinfo {author} {\bibfnamefont {M.~G.}\ \bibnamefont {Raizen}},\
	}\href {\doibase 10.1103/PhysRevLett.67.1735} {\bibfield  {journal} {\bibinfo
			{journal} {Phys. Rev. Lett.}\ }\textbf {\bibinfo {volume} {67}},\ \bibinfo
		{pages} {1735} (\bibinfo {year} {1991})}\BibitemShut {NoStop}%
	\bibitem [{\citenamefont {Heckel}\ \emph {et~al.}(2008)\citenamefont {Heckel},
		\citenamefont {Adelberger}, \citenamefont {Cramer}, \citenamefont {Cook},
		\citenamefont {Schlamminger},\ and\ \citenamefont {Schmidt}}]{Heckel:2008hw}%
	\BibitemOpen
	\bibfield  {author} {\bibinfo {author} {\bibfnamefont {B.~R.}\ \bibnamefont
			{Heckel}}, \bibinfo {author} {\bibfnamefont {E.~G.}\ \bibnamefont
			{Adelberger}}, \bibinfo {author} {\bibfnamefont {C.~E.}\ \bibnamefont
			{Cramer}}, \bibinfo {author} {\bibfnamefont {T.~S.}\ \bibnamefont {Cook}},
		\bibinfo {author} {\bibfnamefont {S.}~\bibnamefont {Schlamminger}}, \ and\
		\bibinfo {author} {\bibfnamefont {U.}~\bibnamefont {Schmidt}},\ }\href
	{\doibase 10.1103/PhysRevD.78.092006} {\bibfield  {journal} {\bibinfo
			{journal} {Phys. Rev. D}\ }\textbf {\bibinfo {volume} {78}},\ \bibinfo
		{pages} {092006} (\bibinfo {year} {2008})},\ \Eprint
	{http://arxiv.org/abs/0808.2673} {arXiv:0808.2673 [hep-ex]} \BibitemShut
	{NoStop}%
	\bibitem [{\citenamefont {Hoedl}\ \emph {et~al.}(2011)\citenamefont {Hoedl},
		\citenamefont {Fleischer}, \citenamefont {Adelberger},\ and\ \citenamefont
		{Heckel}}]{Hoedl:2011zz}%
	\BibitemOpen
	\bibfield  {author} {\bibinfo {author} {\bibfnamefont {S.~A.}\ \bibnamefont
			{Hoedl}}, \bibinfo {author} {\bibfnamefont {F.}~\bibnamefont {Fleischer}},
		\bibinfo {author} {\bibfnamefont {E.~G.}\ \bibnamefont {Adelberger}}, \ and\
		\bibinfo {author} {\bibfnamefont {B.~R.}\ \bibnamefont {Heckel}},\ }\href
	{\doibase 10.1103/PhysRevLett.106.041801} {\bibfield  {journal} {\bibinfo
			{journal} {Phys. Rev. Lett.}\ }\textbf {\bibinfo {volume} {106}},\ \bibinfo
		{pages} {041801} (\bibinfo {year} {2011})}\BibitemShut {NoStop}%
	\bibitem [{\citenamefont {Terrano}\ \emph {et~al.}(2015)\citenamefont
		{Terrano}, \citenamefont {Adelberger}, \citenamefont {Lee},\ and\
		\citenamefont {Heckel}}]{Terrano:2015sna}%
	\BibitemOpen
	\bibfield  {author} {\bibinfo {author} {\bibfnamefont {W.~A.}\ \bibnamefont
			{Terrano}}, \bibinfo {author} {\bibfnamefont {E.~G.}\ \bibnamefont
			{Adelberger}}, \bibinfo {author} {\bibfnamefont {J.~G.}\ \bibnamefont {Lee}},
		\ and\ \bibinfo {author} {\bibfnamefont {B.~R.}\ \bibnamefont {Heckel}},\
	}\href {\doibase 10.1103/PhysRevLett.115.201801} {\bibfield  {journal}
		{\bibinfo  {journal} {Phys. Rev. Lett.}\ }\textbf {\bibinfo {volume} {115}},\
		\bibinfo {pages} {201801} (\bibinfo {year} {2015})},\ \Eprint
	{http://arxiv.org/abs/1508.02463} {arXiv:1508.02463 [hep-ex]} \BibitemShut
	{NoStop}%
	\bibitem [{\citenamefont {Stadnik}\ \emph {et~al.}(2018)\citenamefont
		{Stadnik}, \citenamefont {Dzuba},\ and\ \citenamefont
		{Flambaum}}]{Stadnik:2017hpa}%
	\BibitemOpen
	\bibfield  {author} {\bibinfo {author} {\bibfnamefont {Y.~V.}\ \bibnamefont
			{Stadnik}}, \bibinfo {author} {\bibfnamefont {V.~A.}\ \bibnamefont {Dzuba}},
		\ and\ \bibinfo {author} {\bibfnamefont {V.~V.}\ \bibnamefont {Flambaum}},\
	}\href {\doibase 10.1103/PhysRevLett.120.013202} {\bibfield  {journal}
		{\bibinfo  {journal} {Phys. Rev. Lett.}\ }\textbf {\bibinfo {volume} {120}},\
		\bibinfo {pages} {013202} (\bibinfo {year} {2018})},\ \Eprint
	{http://arxiv.org/abs/1708.00486} {arXiv:1708.00486 [physics.atom-ph]}
	\BibitemShut {NoStop}%
	\bibitem [{\citenamefont {Lee}\ \emph {et~al.}(2018)\citenamefont {Lee},
		\citenamefont {Almasi},\ and\ \citenamefont {Romalis}}]{Lee:2018vaq}%
	\BibitemOpen
	\bibfield  {author} {\bibinfo {author} {\bibfnamefont {J.}~\bibnamefont
			{Lee}}, \bibinfo {author} {\bibfnamefont {A.}~\bibnamefont {Almasi}}, \ and\
		\bibinfo {author} {\bibfnamefont {M.}~\bibnamefont {Romalis}},\ }\href
	{\doibase 10.1103/PhysRevLett.120.161801} {\bibfield  {journal} {\bibinfo
			{journal} {Phys. Rev. Lett.}\ }\textbf {\bibinfo {volume} {120}},\ \bibinfo
		{pages} {161801} (\bibinfo {year} {2018})},\ \Eprint
	{http://arxiv.org/abs/1801.02757} {arXiv:1801.02757 [hep-ex]} \BibitemShut
	{NoStop}%
	\bibitem [{\citenamefont {Dzuba}\ \emph {et~al.}(2018)\citenamefont {Dzuba},
		\citenamefont {Flambaum}, \citenamefont {Samsonov},\ and\ \citenamefont
		{Stadnik}}]{Dzuba:2018anu}%
	\BibitemOpen
	\bibfield  {author} {\bibinfo {author} {\bibfnamefont {V.~A.}\ \bibnamefont
			{Dzuba}}, \bibinfo {author} {\bibfnamefont {V.~V.}\ \bibnamefont {Flambaum}},
		\bibinfo {author} {\bibfnamefont {I.~B.}\ \bibnamefont {Samsonov}}, \ and\
		\bibinfo {author} {\bibfnamefont {Y.~V.}\ \bibnamefont {Stadnik}},\ }\href
	{\doibase 10.1103/PhysRevD.98.035048} {\bibfield  {journal} {\bibinfo
			{journal} {Phys. Rev. D}\ }\textbf {\bibinfo {volume} {98}},\ \bibinfo
		{pages} {035048} (\bibinfo {year} {2018})},\ \Eprint
	{http://arxiv.org/abs/1805.01234} {arXiv:1805.01234 [physics.atom-ph]}
	\BibitemShut {NoStop}%
	\bibitem [{\citenamefont {Fan}\ and\ \citenamefont {Reig}(2023)}]{Fan:2023hci}%
	\BibitemOpen
	\bibfield  {author} {\bibinfo {author} {\bibfnamefont {X.}~\bibnamefont
			{Fan}}\ and\ \bibinfo {author} {\bibfnamefont {M.}~\bibnamefont {Reig}},\
	}\href@noop {} {\  (\bibinfo {year} {2023})},\ \Eprint
	{http://arxiv.org/abs/2310.18797} {arXiv:2310.18797 [hep-ph]} \BibitemShut
	{NoStop}%
	\bibitem [{\citenamefont {Agrawal}\ \emph
		{et~al.}(2023{\natexlab{a}})\citenamefont {Agrawal}, \citenamefont {Hutzler},
		\citenamefont {Kaplan}, \citenamefont {Rajendran},\ and\ \citenamefont
		{Reig}}]{Agrawal:2023lmw}%
	\BibitemOpen
	\bibfield  {author} {\bibinfo {author} {\bibfnamefont {P.}~\bibnamefont
			{Agrawal}}, \bibinfo {author} {\bibfnamefont {N.~R.}\ \bibnamefont
			{Hutzler}}, \bibinfo {author} {\bibfnamefont {D.~E.}\ \bibnamefont {Kaplan}},
		\bibinfo {author} {\bibfnamefont {S.}~\bibnamefont {Rajendran}}, \ and\
		\bibinfo {author} {\bibfnamefont {M.}~\bibnamefont {Reig}},\ }\href@noop {}
	{\  (\bibinfo {year} {2023}{\natexlab{a}})},\ \Eprint
	{http://arxiv.org/abs/2309.10023} {arXiv:2309.10023 [hep-ph]} \BibitemShut
	{NoStop}%
	\bibitem [{\citenamefont {Baruch}\ \emph
		{et~al.}(2024{\natexlab{a}})\citenamefont {Baruch}, \citenamefont {Changala},
		\citenamefont {Shagam},\ and\ \citenamefont {Soreq}}]{Baruch:2024frj}%
	\BibitemOpen
	\bibfield  {author} {\bibinfo {author} {\bibfnamefont {C.}~\bibnamefont
			{Baruch}}, \bibinfo {author} {\bibfnamefont {P.~B.}\ \bibnamefont
			{Changala}}, \bibinfo {author} {\bibfnamefont {Y.}~\bibnamefont {Shagam}}, \
		and\ \bibinfo {author} {\bibfnamefont {Y.}~\bibnamefont {Soreq}},\
	}\href@noop {} {\  (\bibinfo {year} {2024}{\natexlab{a}})},\ \Eprint
	{http://arxiv.org/abs/2402.07504} {arXiv:2402.07504 [hep-ph]} \BibitemShut
	{NoStop}%
	\bibitem [{\citenamefont {Baruch}\ \emph
		{et~al.}(2024{\natexlab{b}})\citenamefont {Baruch}, \citenamefont {Changala},
		\citenamefont {Shagam},\ and\ \citenamefont {Soreq}}]{Baruch:2024fbh}%
	\BibitemOpen
	\bibfield  {author} {\bibinfo {author} {\bibfnamefont {C.}~\bibnamefont
			{Baruch}}, \bibinfo {author} {\bibfnamefont {P.~B.}\ \bibnamefont
			{Changala}}, \bibinfo {author} {\bibfnamefont {Y.}~\bibnamefont {Shagam}}, \
		and\ \bibinfo {author} {\bibfnamefont {Y.}~\bibnamefont {Soreq}},\
	}\href@noop {} {\  (\bibinfo {year} {2024}{\natexlab{b}})},\ \Eprint
	{http://arxiv.org/abs/2406.02281} {arXiv:2406.02281 [hep-ph]} \BibitemShut
	{NoStop}%
	\bibitem [{\citenamefont {Wei}\ \emph {et~al.}(2023)\citenamefont {Wei},
		\citenamefont {Zhao}, \citenamefont {Fang}, \citenamefont {Xu}, \citenamefont
		{Liu}, \citenamefont {Cao}, \citenamefont {Wickenbrock}, \citenamefont {Hu},
		\citenamefont {Ji},\ and\ \citenamefont {Budker}}]{Wei:2022ggs}%
	\BibitemOpen
	\bibfield  {author} {\bibinfo {author} {\bibfnamefont {K.}~\bibnamefont
			{Wei}}, \bibinfo {author} {\bibfnamefont {T.}~\bibnamefont {Zhao}}, \bibinfo
		{author} {\bibfnamefont {X.}~\bibnamefont {Fang}}, \bibinfo {author}
		{\bibfnamefont {Z.}~\bibnamefont {Xu}}, \bibinfo {author} {\bibfnamefont
			{C.}~\bibnamefont {Liu}}, \bibinfo {author} {\bibfnamefont {Q.}~\bibnamefont
			{Cao}}, \bibinfo {author} {\bibfnamefont {A.}~\bibnamefont {Wickenbrock}},
		\bibinfo {author} {\bibfnamefont {Y.}~\bibnamefont {Hu}}, \bibinfo {author}
		{\bibfnamefont {W.}~\bibnamefont {Ji}}, \ and\ \bibinfo {author}
		{\bibfnamefont {D.}~\bibnamefont {Budker}},\ }\href {\doibase
		10.1103/PhysRevLett.130.063201} {\bibfield  {journal} {\bibinfo  {journal}
			{Phys. Rev. Lett.}\ }\textbf {\bibinfo {volume} {130}},\ \bibinfo {pages}
		{063201} (\bibinfo {year} {2023})},\ \Eprint
	{http://arxiv.org/abs/2210.09027} {arXiv:2210.09027 [physics.atom-ph]}
	\BibitemShut {NoStop}%
	\bibitem [{\citenamefont {Agrawal}\ \emph
		{et~al.}(2023{\natexlab{b}})\citenamefont {Agrawal}, \citenamefont {Kaplan},
		\citenamefont {Kim}, \citenamefont {Rajendran},\ and\ \citenamefont
		{Reig}}]{Agrawal:2022wjm}%
	\BibitemOpen
	\bibfield  {author} {\bibinfo {author} {\bibfnamefont {P.}~\bibnamefont
			{Agrawal}}, \bibinfo {author} {\bibfnamefont {D.~E.}\ \bibnamefont {Kaplan}},
		\bibinfo {author} {\bibfnamefont {O.}~\bibnamefont {Kim}}, \bibinfo {author}
		{\bibfnamefont {S.}~\bibnamefont {Rajendran}}, \ and\ \bibinfo {author}
		{\bibfnamefont {M.}~\bibnamefont {Reig}},\ }\href {\doibase
		10.1103/PhysRevD.108.015017} {\bibfield  {journal} {\bibinfo  {journal}
			{Phys. Rev. D}\ }\textbf {\bibinfo {volume} {108}},\ \bibinfo {pages}
		{015017} (\bibinfo {year} {2023}{\natexlab{b}})},\ \Eprint
	{http://arxiv.org/abs/2210.17547} {arXiv:2210.17547 [hep-ph]} \BibitemShut
	{NoStop}%
	\bibitem [{\citenamefont {Scherer}\ and\ \citenamefont
		{Schindler}(2012)}]{Scherer:2012xha}%
	\BibitemOpen
	\bibfield  {author} {\bibinfo {author} {\bibfnamefont {S.}~\bibnamefont
			{Scherer}}\ and\ \bibinfo {author} {\bibfnamefont {M.~R.}\ \bibnamefont
			{Schindler}},\ }\href {\doibase 10.1007/978-3-642-19254-8} {\emph {\bibinfo
			{title} {{A Primer for Chiral Perturbation Theory}}}},\ Vol.\ \bibinfo
	{volume} {830}\ (\bibinfo  {publisher} {Springer Berlin, Heidelberg},\
	\bibinfo {year} {2012})\BibitemShut {NoStop}%
	\bibitem [{\citenamefont {Hoferichter}\ \emph {et~al.}(2015)\citenamefont
		{Hoferichter}, \citenamefont {Ruiz~de Elvira}, \citenamefont {Kubis},\ and\
		\citenamefont {Mei\ss{}ner}}]{Hoferichter:2015dsa}%
	\BibitemOpen
	\bibfield  {author} {\bibinfo {author} {\bibfnamefont {M.}~\bibnamefont
			{Hoferichter}}, \bibinfo {author} {\bibfnamefont {J.}~\bibnamefont {Ruiz~de
				Elvira}}, \bibinfo {author} {\bibfnamefont {B.}~\bibnamefont {Kubis}}, \ and\
		\bibinfo {author} {\bibfnamefont {U.-G.}\ \bibnamefont {Mei\ss{}ner}},\
	}\href {\doibase 10.1103/PhysRevLett.115.092301} {\bibfield  {journal}
		{\bibinfo  {journal} {Phys. Rev. Lett.}\ }\textbf {\bibinfo {volume} {115}},\
		\bibinfo {pages} {092301} (\bibinfo {year} {2015})},\ \Eprint
	{http://arxiv.org/abs/1506.04142} {arXiv:1506.04142 [hep-ph]} \BibitemShut
	{NoStop}%
	\bibitem [{\citenamefont {Hoferichter}\ \emph {et~al.}(2016)\citenamefont
		{Hoferichter}, \citenamefont {Ruiz~de Elvira}, \citenamefont {Kubis},\ and\
		\citenamefont {Mei\ss{}ner}}]{Hoferichter:2016ocj}%
	\BibitemOpen
	\bibfield  {author} {\bibinfo {author} {\bibfnamefont {M.}~\bibnamefont
			{Hoferichter}}, \bibinfo {author} {\bibfnamefont {J.}~\bibnamefont {Ruiz~de
				Elvira}}, \bibinfo {author} {\bibfnamefont {B.}~\bibnamefont {Kubis}}, \ and\
		\bibinfo {author} {\bibfnamefont {U.-G.}\ \bibnamefont {Mei\ss{}ner}},\
	}\href {\doibase 10.1016/j.physletb.2016.06.038} {\bibfield  {journal}
		{\bibinfo  {journal} {Phys. Lett. B}\ }\textbf {\bibinfo {volume} {760}},\
		\bibinfo {pages} {74} (\bibinfo {year} {2016})},\ \Eprint
	{http://arxiv.org/abs/1602.07688} {arXiv:1602.07688 [hep-lat]} \BibitemShut
	{NoStop}%
	\bibitem [{\citenamefont {An}\ \emph {et~al.}(2010)\citenamefont {An},
		\citenamefont {Ji},\ and\ \citenamefont {Xu}}]{An:2009zh}%
	\BibitemOpen
	\bibfield  {author} {\bibinfo {author} {\bibfnamefont {H.}~\bibnamefont
			{An}}, \bibinfo {author} {\bibfnamefont {X.}~\bibnamefont {Ji}}, \ and\
		\bibinfo {author} {\bibfnamefont {F.}~\bibnamefont {Xu}},\ }\href {\doibase
		10.1007/JHEP02(2010)043} {\bibfield  {journal} {\bibinfo  {journal} {JHEP}\
		}\textbf {\bibinfo {volume} {02}},\ \bibinfo {pages} {043} (\bibinfo {year}
		{2010})},\ \Eprint {http://arxiv.org/abs/0908.2420} {arXiv:0908.2420
		[hep-ph]} \BibitemShut {NoStop}%
	\bibitem [{\citenamefont {Guo}\ and\ \citenamefont
		{Meissner}(2012)}]{Guo:2012vf}%
	\BibitemOpen
	\bibfield  {author} {\bibinfo {author} {\bibfnamefont {F.-K.}\ \bibnamefont
			{Guo}}\ and\ \bibinfo {author} {\bibfnamefont {U.-G.}\ \bibnamefont
			{Meissner}},\ }\href {\doibase 10.1007/JHEP12(2012)097} {\bibfield  {journal}
		{\bibinfo  {journal} {JHEP}\ }\textbf {\bibinfo {volume} {12}},\ \bibinfo
		{pages} {097} (\bibinfo {year} {2012})},\ \Eprint
	{http://arxiv.org/abs/1210.5887} {arXiv:1210.5887 [hep-ph]} \BibitemShut
	{NoStop}%
\end{thebibliography}
\end{document}